\documentclass[5p,twocolumn,number,sort&compress]{elsarticle}

\newcommand{\Vec}[1]{\mbox{\boldmath $#1$}}

\title{Coupled map lattice for the spiral pattern formation in astronomical objects}
\author{Erika Nozawa}

\begin{document}

%\tnotetext[t1]{Declarations of interest: none}
\ead{pop2019@e-rika.net}
\address{Department of Physics, Graduate School of Humanities and Sciences, Ochanomizu University,\\
2-1-1 Ohtsuka, Bunkyo-ku, Tokyo 112-8610, Japan}
\date{January 28, 2020}
\begin{abstract}
We propose a minimal coupled map lattice (CML)
for the spiral pattern formation in astronomical objects which consist of accreting gas
induced by gravity as a long-range force.
In the proposed CML, we assume only two procedures:
one in which the flow of gas particles occurs by gravity
and another one in which the collision of gas particles occurs by advection in the flow.
In spite of its simplicity,
the numerical simulation of the proposed CML shows a new formation process
in which grand design spiral patterns appear due to gas ejection from a central star.
Several aspects of the formation process are indeed in good agreement with the results of conventional theories and observations.
This agreement and the observations of outflows should suggest that the above gas ejection can happen and lead to the formation of grand design spiral patterns, in actual spiral galaxies and protoplanetary disks.
\end{abstract}
\maketitle

\section{Introduction}

Coupled map lattice (CML) reproduces some important features of
spatially extended dynamical phenomena\cite{Kaneko,Kanekos}.
In other words, CML simulates dynamic patterns and properties well which appear in such phenomena.
CML is a type of dynamical systems with discrete space and time, and continuous state variables.
As representative examples of such CML,
the CMLs for boiling\cite{Yanagitab}, convection\cite{Yanagitac}, cloud formation\cite{Yanagitad}
and sand ripple formation\cite{Nishimori} have been proposed.
%One advantage of CML lies in simple construction of time evolution by combining procedures,
%as described in terms of the "constructive approach"\cite{Kaneko}.
In the CML for a dynamical phenomenon,
the time evolution of the field variables on the lattice is given by successive operations of procedures.
Here the procedures are maps acting on the field variables
and derived from decomposing the dynamical phenomenon into important elementary processes.
Thus, in approaches to constructing time evolution, CML is different from conventional computation, such as setting up the partial differential equations for a fluid system at first and discretizing them properly.
This CML approach is quite useful to find the essential elementary processes in dynamical phenomena.
Indeed, many CMLs including the above examples\cite{Yanagitab,Yanagitac,Yanagitad,Nishimori} have been reported,
in which the dynamic behavior
is in good agreement with the experimental observations of the corresponding phenomenon.

In this paper, we introduce a minimal CML
which shows the spiral pattern formation in astronomical objects consisting of accreting cold dense gas\cite{GalaDy},
such as seen in spiral galaxies\cite{GalaDy} or protoplanetary disks\cite{SEEDS}.
It is constructed by a minimal set of procedures for simulating such pattern formation:
a procedure in which the flow of gas particles occurs by gravity
and another one in which the collision of gas particles occurs by advection in the flow.
It is different from
the above examples\cite{Yanagitab,Yanagitac,Yanagitad,Nishimori} in that long-range force (gravity) acts.
We expect that this minimal set of procedures should reveal to us necessary (and therefore essential) elementary processes in the pattern formation,
as argued in terms of the "reductionism in procedure"\cite{Kaneko}.

The proposed CML shows a new formation process
in which grand design spiral patterns appear due to gas ejection from a central star.
Starting from a random initial state,
gas clumps (that is, macroscopic clumps of gas particles) gravitate each other to gather around the center of the lattice
and form a central star consisting of four massive gas clumps.
The central star then contracts, ejects gas particles and then begins to expand.
The ejected gas particles jam Keplerian gas particles around the central star.
The jammed Keplerian gas particles form a pair of spiral arms and thus a grand design spiral pattern (that is, two-arm spiral pattern) appears.
After that, the central star begins to contract again, the spiral pattern becomes less sharp
and then the second gas ejection from the central star occurs.
The above pattern formation and disappearance are repeated over and over again for a long time.

Three aspects of the formation process are indeed in good agreement with the results of the conventional theories and observations\cite{GalaDy,Lin-Shu,Kuno,Miyoshi,Nadia,Kuno2}.
First, the simulated spiral patterns are grand design spirals with two gaseous spiral arms, which agrees with the observations in spiral galaxies\cite{Kuno}.
Second, gas particles are in Keplerian motion around a central star, which is a fundamental dynamic property of astronomical objects\cite{GalaDy} and agrees with the observations\cite{Miyoshi,Nadia}.
Third, spiral arms are formed by jammed Keplerian gas particles, which is a property of density waves\cite{Lin-Shu} and agrees with the observations\cite{Kuno2}.
This agreement and also the observations of outflows\cite{Tombesi,Matsushita} should suggest that the above gas ejection can happen and lead to the formation of grand design spiral patterns, in actual spiral galaxies and protoplanetary disks.

We have shown the formation of diverse patterns in astronomical objects
by the qualitative and fast computational method of the proposed CML,
after searching a wide range of the parameters and initial conditions.
This fast computational method could give us insightful suggestions on the formation mechanism of astronomical objects including undiscovered phenomena, as described above.
It has different advantages to the quantitative but slow computational method of adaptive mesh refinement (AMR) \cite{AMRo,Matsumoto} and smoothed particle hydrodynamics (SPH) \cite{SPHo,Springel}
both of which come into their own for detailed comparisons with observations.
The simulations in AMR or SPH are usually carried out by supercomputers.
In contrast, the simulations in the proposed CML are carried out enough
by a personal computer\footnote{Indeed, it takes less than 2 minutes (113 seconds)
to obtain the simulation results to be shown as Fig.\,\ref{fig:alpha_0.eps}.
It is also emphasized that
I have developed the simulator of the proposed CML in low-execution-speed environment consisting of JavaScript and HTML.}
(note that the proposed CML is suitable for parallel computing well).

Moreover, the proposed CML performs stable and fast computation since it is constructed flexibly
by choosing the Eulerian procedure\cite{Kaneko,Yanagitac} or the Lagrangian procedure\cite{Kaneko,Yanagitac} properly.
The Eulerian method such as AMR uses grids (meshes),
which has the advantage of fast computation rather than stable computation.
The Lagrangian method such as SPH uses particles,
which has the advantage of stable computation rather than fast computation.
Thus the proposed CML has the advantages of both the Eulerian and Lagrangian methods.

The present paper is organized as follows.
In section \ref{model},
we construct a CML which consists of gravitational interaction and advection procedures.
In section \ref{simulation},
we show the snapshots of the simulations and explain how diverse patterns are dynamically formed,
especially focusing on a new formation process of grand design spiral patterns.
In section \ref{Keplerian motion of gas clumps},
we verify that gas clumps are in Keplerian motion around a central star
by using the snapshot of a velocity field of gas clumps and the rotation curve of a spiral pattern.
Summary and discussion are given in Section~\ref{summary}.

\section{Model}
\label{model}

\begin{figure*}[t]
  \begin{center}
    \includegraphics[scale=0.5]{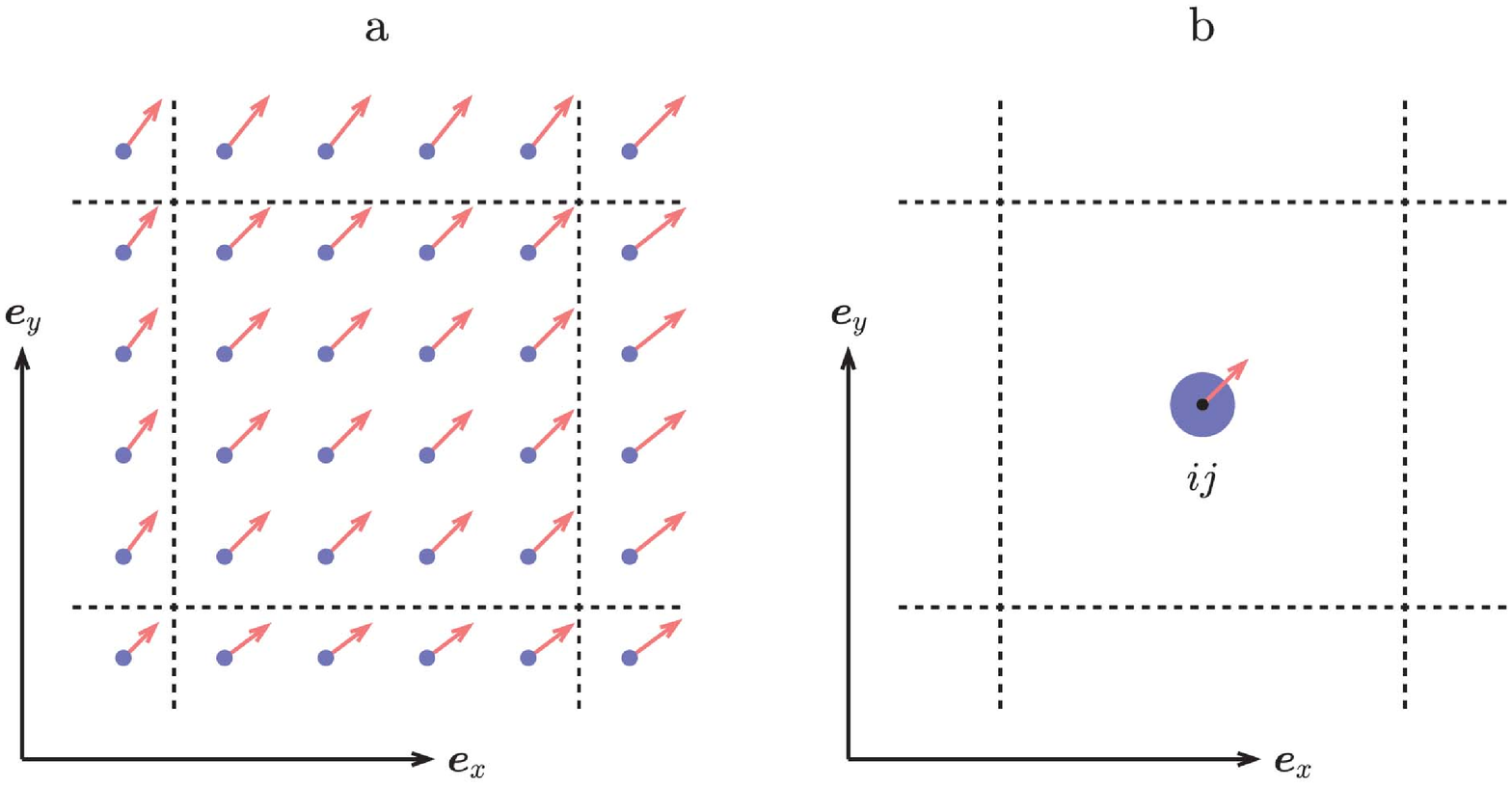}
  \end{center}
  \caption{
	Two complementary pictures for the field variables.
  (a) Particle picture. The black dotted square represents the cell at the lattice point $ij$.
  In the cell, the set of the blue dots represents the total mass of gas particles (gas clump mass $m_{ij}^{t}$)
  and the set of the red arrows the flow of them (gas clump velocity $\Vec{v}_{ij}^{t}$).
  (b) Lattice picture. The black dot represents the lattice point $ij$.
  At the lattice point $ij$, the blue filled circle represents the mass of the gas clump (gas clump mass $m_{ij}^{t}$)
  and the red arrow the velocity of it (gas clump velocity $\Vec{v}_{ij}^{t}$).
  }
  \label{fig:fv.eps}
\end{figure*}

Let us construct a coupled map lattice (CML)
for the spiral pattern formation in astronomical objects consisting of accreting gas,
making use of the CML modeling algorithm\cite{Kaneko,Kanekos}.

\subsection{System and lattice}

We consider a system in which accreting cold dense gas is moving on a two-dimensional region in three-dimensional space,
such as an accretion disk\cite{GalaDy}.
We divide the region into $N_{x}\times N_{y}$ square cells.
For simplicity, in each cell,
we treat the gas as an assembly of virtual gas particles (not gas molecules)
which are distributed uniformly, carried by the same flow and have the same velocity (see the particle picture in Fig.\,\ref{fig:fv.eps}a).
From another viewpoint, there is a macroscopic gas clump formed by the collision of dense gas particles in each cell.

We introduce a finite two-dimensional square lattice by locating a lattice point on the center of each cell.
Hereafter we set the distance between the nearest neighbor lattice points to one
(thus, the size of each cell is one),
label the lattice points with $ij$ ($i=0,1,\cdots,N_{x}-1$ and $j=0,1,\cdots,N_{y}-1$)
and express their positions by the position vectors $\Vec{r}_{ij}=(i,j)=i\Vec{e}_{x}+j\Vec{e}_{y}$,
where $\Vec{e}_{x}$ and $\Vec{e}_{y}$ are unit vectors parallel to the $x$-axis and $y$-axis respectively.

\subsection{Field variables}

We prepare a set of macroscopically coarse-grained field variables on the lattice.
As the field variables at lattice point $ij$ at discrete time $t$,
we define the gas clump mass $m_{ij}^{t}$
and the gas clump velocity $\Vec{v}_{ij}^{t}=v_{x\, ij}^{t}\Vec{e}_{x}+v_{y\, ij}^{t}\Vec{e}_{y}$.

There are two complementary pictures for the field variables: the lattice and particle pictures (see Fig.\,\ref{fig:fv.eps}).
In the lattice picture (see Fig.\,\ref{fig:fv.eps}b),
the gas clump mass $m_{ij}^{t}$ and velocity $\Vec{v}_{ij}^{t}$ represent
the mass (the blue filled circle in Fig.\,\ref{fig:fv.eps}b) and velocity (the red arrow in Fig.\,\ref{fig:fv.eps}b)
of the gas clump at lattice point $ij$ (the black dot in Fig.\,\ref{fig:fv.eps}b), respectively.
On the other hand, in the particle picture (see Fig.\,\ref{fig:fv.eps}a),
they represent the total mass (the set of the blue dots in Fig.\,\ref{fig:fv.eps}a)
and the flow (the set of the red arrows in Fig.\,\ref{fig:fv.eps}a) of gas particles in the cell at lattice point $ij$, respectively.

\subsection{Elementary processes}

The spiral pattern formation in gaseous astronomical objects
is decomposed into the following two important elementary processes:
a gravitational interaction process and an advection process.
In the former process,
the gravitational interaction among gas clumps
leads each gas clump to a change in the gas clump velocity (that is, the flow of gas particles in the gas clump).
In the latter process,
the flows resulting from the gravitational interaction
make gas particles move and collide with their mass and momentum, and form new gas clumps.

\subsection{Procedures}

\subsubsection{Gravitational interaction procedure}

We formulate the gravitational interaction process as an Eulerian procedure $T_{g}$ in the lattice picture.
Here Eulerian procedures generally describe the change in field variables through the interaction among them\cite{Kaneko,Yanagitac}.
In the procedure $T_{g}$,
the gas clump at lattice point $ij$ (of gas clump mass $m_{ij}^{t}$) is given an impulse
by the gravitational interaction from the other gas clumps at lattice points $kl$ (of gas clump masses $m_{kl}^{t}$)
and then changes the gas clump velocity $\Vec{v}_{ij}^{t}$ to $\Vec{v}_{ij}^{*}$,
where $*$ represents an intermediate time between discrete times $t$ and $t+1$.
Thus, the gravitational interaction procedure $T_{g}$ is defined by the following maps:
\begin{equation}
\label{eqn:Tg_m}
m_{ij}^{*}=m_{ij}^{t},
\end{equation}
\begin{eqnarray}
\label{eqn:Tg_v}
\lefteqn{\Vec{v}_{ij}^{*}=\Vec{v}_{ij}^{t}}
\hphantom{-\gamma\tau_{g}\sum_{k=0}^{N_{\mathstrut x}-1}\sum_{l=0}^{N_{\mathstrut y}-1}
\frac{(1-\delta_{ik}\delta_{jl})m_{kl}^{t}}{|\Vec{r}_{ij}-\Vec{r}_{kl}|^{2}}
\frac{\Vec{r}_{ij}-\Vec{r}_{kl}}{|\Vec{r}_{ij}-\Vec{r}_{kl}|},}\nonumber\\
-\gamma\tau_{g}\sum_{k=0}^{N_{\mathstrut x}-1}\sum_{l=0}^{N_{\mathstrut y}-1}
\frac{(1-\delta_{ik}\delta_{jl})m_{kl}^{t}}{|\Vec{r}_{ij}-\Vec{r}_{kl}|^{2}}
\frac{\Vec{r}_{ij}-\Vec{r}_{kl}}{|\Vec{r}_{ij}-\Vec{r}_{kl}|},
\end{eqnarray}
where $\gamma$ is the gravitational constant,
$\tau_{g}$ the time interval for the procedure $T_{g}$
and $\delta$ the Kronecker delta.
As shown in Eq.\,\ref{eqn:Tg_m},
the gas clump mass $m_{ij}^{t}$ does not change in the procedure $T_{g}$.

\subsubsection{Advection procedure}

We formulate the advection process as a Lagrangian procedure $T_{a}$ in the particle picture.
Here Lagrangian procedures generally describe the change in field variables along the flow of particles\cite{Kaneko,Yanagitac}.
In the procedure $T_{a}$,
each flow $\Vec{v}_{kl}^{*}$ resulting from the procedure $T_{g}$
carries gas particles with their total mass $m_{kl}^{*}$ and momentum $m_{kl}^{*}\Vec{v}_{kl}^{*}$
from the cell at lattice point $kl$
to a cell-sized area centered at the position
\begin{equation}
\label{eqn:kl}
(\tilde{k},\tilde{l})=(k+v_{x\, kl}^{*}\tau_{a},l+v_{y\, kl}^{*}\tau_{a}),
\end{equation}
where $\tau_{a}$ is the time interval for the procedure $T_{a}$
(see Fig.\,\ref{fig:advection.eps}b in \ref{Advection procedure}).
When the cell-sized areas overlap the cell at lattice point $ij$,
the size of each overlap area is given by
\begin{eqnarray}
\label{eqn:w_ijkl}
\lefteqn{w_{ijkl}^{*}=
\left(\delta_{i\lfloor \tilde{k}\rfloor}\delta_{j\lfloor \tilde{l}\rfloor}
+\delta_{i\lfloor \tilde{k}\rfloor +1}\delta_{j\lfloor \tilde{l}\rfloor}
+\delta_{i\lfloor \tilde{k}\rfloor +1}\delta_{j\lfloor \tilde{l}\rfloor +1}\right.}
\hphantom{\left.
+\delta_{i\lfloor \tilde{k}\rfloor}\delta_{j\lfloor \tilde{l}\rfloor +1}\right)
\left(1-\left|\tilde{k}-i\right|\right)\left(1-\left|\tilde{l}-j\right|\right),}\nonumber\\
\left.
+\delta_{i\lfloor \tilde{k}\rfloor}\delta_{j\lfloor \tilde{l}\rfloor +1}\right)
\left(1-\left|\tilde{k}-i\right|\right)\left(1-\left|\tilde{l}-j\right|\right),
\end{eqnarray}
where $\lfloor \bullet \rfloor$ is the floor function.
In Eq.\,\ref{eqn:w_ijkl}, the first brackets give a flag to become one
when the overlapping occurs or zero when does not,
and the second and third the size when the first brackets become one
(for details, see \ref{Advection procedure}).
With Eq.\,\ref{eqn:w_ijkl}, in each overlap area,
the total mass and momentum of gas particles are given by
$w_{ijkl}^{*}m_{kl}^{*}$ and $w_{ijkl}^{*}m_{kl}^{*}\Vec{v}_{kl}^{*}$ respectively.
In the cell at lattice point $ij$,
gas particles in the overlap areas collide with each other and are mixed into one.
Through this collision and mixture,
they form a new gas clump whose mass and velocity are $m_{ij}^{t+1}$ and $\Vec{v}_{ij}^{t+1}$ respectively.
Thus, the advection procedure $T_{a}$ is defined by the following maps:
\begin{equation}
\label{eqn:Ta_m}
m_{ij}^{t+1}=\sum_{k=0}^{N_{\mathstrut x}-1}\sum_{l=0}^{N_{\mathstrut y}-1}w_{ijkl}^{*}m_{kl}^{*},
\end{equation}
\begin{equation}
\label{eqn:Ta_v}
\Vec{v}_{ij}^{t+1}=\frac{1}{m_{ij}^{t+1}}\sum_{k=0}^{N_{\mathstrut x}-1}\sum_{l=0}^{N_{\mathstrut y}-1}
w_{ijkl}^{*}m_{kl}^{*}\Vec{v}_{kl}^{*}.
\end{equation}
In Eq.\,\ref{eqn:Ta_v}, when gas clump mass $m_{ij}^{t+1}$ takes zero,
gas clump velocity $\Vec{v}_{ij}^{t+1}$ is set to also zero.
We may operate the procedure $T_{a}$ of Eqs.\,\ref{eqn:Ta_m} and \ref{eqn:Ta_v}
with a low computational cost of $O(N)$, where $N$ is the total number of lattice points
(for details, see \ref{Computational cost of advection procedure}).

\subsection{Time evolution}

By successive operations of the gravitational interaction procedure $T_{g}$ and the advection procedure $T_{a}$,
the time evolution of gas clump mass $m_{ij}^{t}$ and velocity $\Vec{v}_{ij}^{t}$
by one step (from discrete time $t$ to $t+1$) is constructed as follows:
\begin{equation}
\label{eqn:dynamics}
\left(
\begin{array}{c}
{m_{ij}^{t}\rule{0pt}{10pt}} \\
{\Vec{v}_{ij}^{t}\rule{0pt}{10pt}} \\
\end{array}
\right)
\stackrel{T_{g}}{\longmapsto}
\left(
\begin{array}{c}
{m_{ij}^{*}\rule{0pt}{10pt}} \\
{\Vec{v}_{ij}^{*}\rule{0pt}{10pt}} \\
\end{array}
\right)
\stackrel{T_{a}}{\longmapsto}
\left(
\begin{array}{c}
{m_{ij}^{t+1}\rule{0pt}{10pt}} \\
{\Vec{v}_{ij}^{t+1}\rule{0pt}{10pt}} \\
\end{array}
\right).
\end{equation}
On the time evolution of Eq.\,\ref{eqn:dynamics},
the total mass $m_{G}^{t}$, total momentum $\Vec{p}_{G}^{t}$ and total angular momentum $\Vec{l}_{G}^{t}$ of the system are conserved
(for details, see \ref{Conservative quantities}).
These constant quantities are given by
\begin{equation}
m_{G}^{t}=\sum_{i=0}^{N_{\mathstrut x}-1}\sum_{j=0}^{N_{\mathstrut y}-1}m_{ij}^{t},
\end{equation}
\begin{equation}
\Vec{p}_{G}^{t}=\sum_{i=0}^{N_{\mathstrut x}-1}\sum_{j=0}^{N_{\mathstrut y}-1}m_{ij}^{t}\Vec{v}_{ij}^{t},
\end{equation}
\begin{equation}
\Vec{l}_{G}^{t}=\sum_{i=0}^{N_{\mathstrut x}-1}\sum_{j=0}^{N_{\mathstrut y}-1}
\left(\Vec{r}_{ij}-\Vec{r}_{G}^{t}\right)\times m_{ij}^{t}\Vec{v}_{ij}^{t},
\end{equation}
where $\Vec{r}_{G}^{t}$ is the position vector of the center of gravity of the system, given by
\begin{equation}
\Vec{r}_{G}^{t}=\frac{1}{m_{G}^{t}}\sum_{i=0}^{N_{\mathstrut x}-1}\sum_{j=0}^{N_{\mathstrut y}-1}m_{ij}^{t}\Vec{r}_{ij}.
\end{equation}

The simulations are performed according to the following settings:
Lattice size $N_{x}\times N_{y}$ is $50\times 50$;
Gravitational constant $\gamma$ is one and time intervals $\tau_{g}$ and $\tau_{a}$ one;
The initial gas clump mass $m_{ij}^{0}$ is given by a uniform random number within $[0,2\mu/(N_{x}N_{y})]$
and initial gas clump velocity $\Vec{v}_{ij}^{0}$ zero;
The boundary conditions are open.
Under the above initial conditions,
the initial total mass $m_{G}^{0}$ approximately has the normal distribution with the expected value $\mu$ and the variance $\sigma^{2}=\mu^{2}/(3N_{x}N_{y})$.

In the simulations, the total mass $m_{G}^{t}$, total momentum $\Vec{p}_{G}^{t}$ and total angular momentum $\Vec{l}_{G}^{t}$ are not conserved completely,
since gas particles move out through the boundary of the finite lattice.
We note that their mass is a very small quantity ($0.7\%$ of the initial total mass)
through the simulation shown in Fig.\,\ref{fig:alpha_0.eps}.

\section{Dynamic formation of diverse patterns}
\label{simulation}

\begin{figure*}[t]
  \begin{center}
    \includegraphics[scale=1.0]{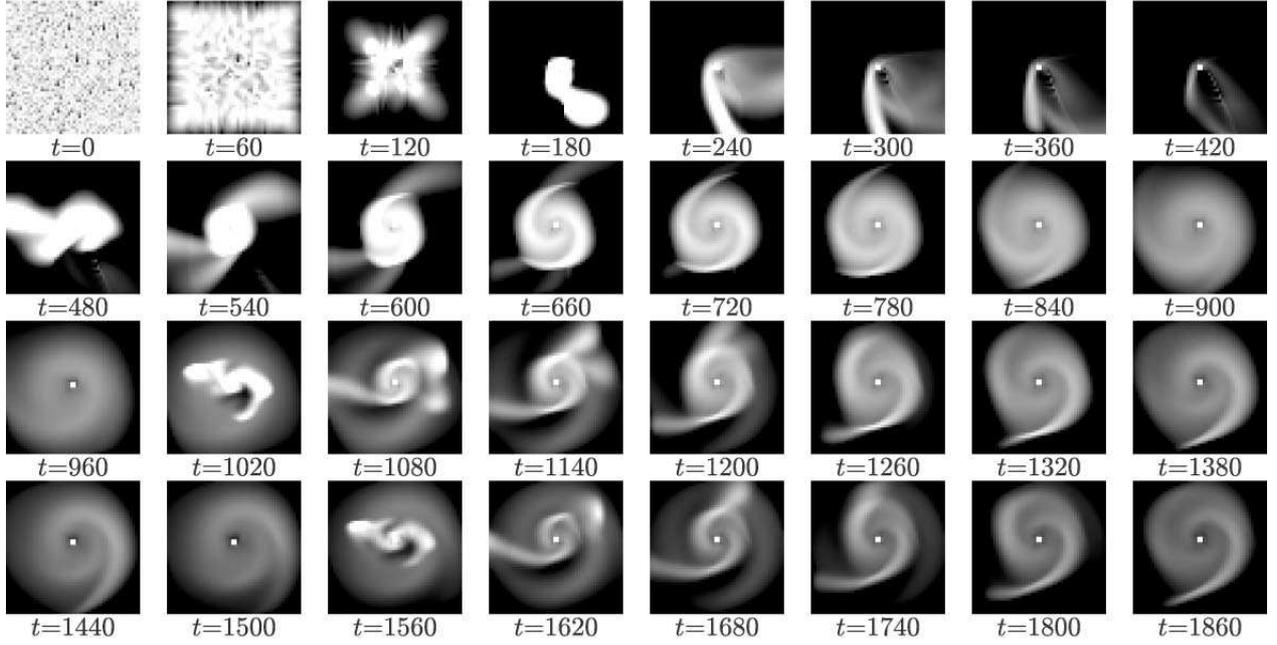}
  \end{center}
  \caption{
	Snapshots of gas clump masses $m_{ij}^{t}$.
  The brightness of cells represents the logarithm of gas clump masses $\log_{10}m_{ij}^{t}$
  ($i=0,1,\cdots,49$, $j=0,1,\cdots,49$ and $t=0,60,\cdots,1860$)
  in the range of $1\times 10^{-7}$ to $8\times 10^{-4}$ at $t=0$,
  $1\times 10^{-7}$ to $5.01\times 10^{-4}$ at $t=60$, $1\times 10^{-7}$ to $2.02\times 10^{-4}$ at $t=120$
  and $1\times 10^{-7}$ to $3\times 10^{-6}$ at $t=180,240,\cdots,1860$.
  }
  \label{fig:alpha_0.eps}
\end{figure*}

\begin{figure*}[t]
  \begin{center}
    \includegraphics[scale=1.0]{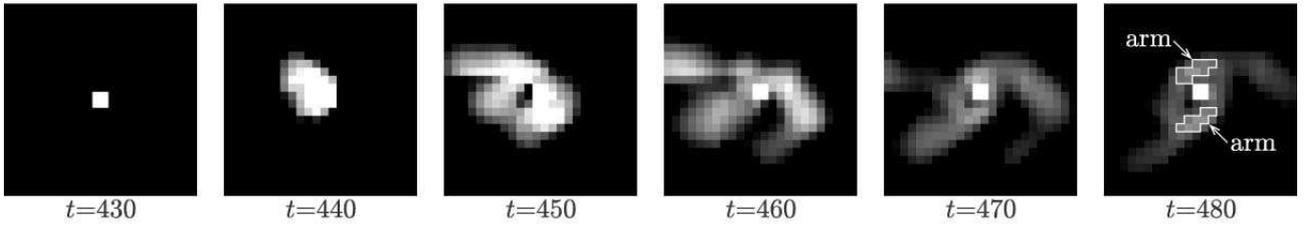}
  \end{center}
  \caption{
	Enlarged snapshots of gas clump masses $m_{ij}^{t}$ around the central star.
  The brightness of cells represents the logarithm of gas clump masses $\log_{10}m_{ij}^{t}$
  ($i=13,14,\cdots,36$, $j=13,14,\cdots,36$ and $t=430,440,\cdots,480$)
  in the range of $6\times 10^{-6}$ to $6\times 10^{-5}$.
 	Two newborn arms are surrounded by white lines.
  }
  \label{fig:emf.eps}
\end{figure*}

Let us see how diverse patterns, especially grand design spiral patterns, are dynamically formed in the simulation of the proposed CML
by using the snapshots of gas clump masses $m_{ij}^{t}$ shown in Figs.\,\ref{fig:pd.eps} and \ref{fig:alpha_0.eps}, and the enlarged ones in Fig.\,\ref{fig:emf.eps}.
Each snapshot has $50\times 50$ cells in Figs.\,\ref{fig:pd.eps} and \ref{fig:alpha_0.eps}, and $24\times 24$ cells in Fig.\,\ref{fig:emf.eps}.
The brightness of the cells represents the logarithm of gas clump masses $\log_{10}m_{ij}^{t}$.

\subsection{Dependence of patterns on the initial total mass}
\label{Dependence of patterns on the initial total mass}

\begin{figure}[htbp]
  \begin{center}
    \includegraphics[scale=1.0]{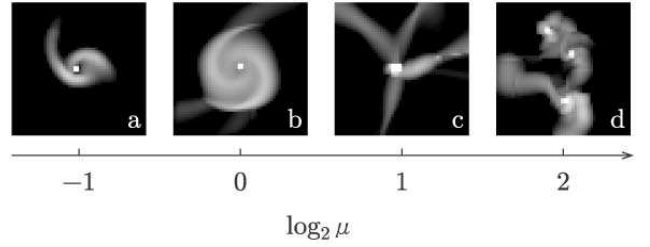}
  \end{center}
  \caption{
	Snapshots of typical patterns with different expected values of initial total mass $\mu$.
  The brightness of cells represents the logarithm of gas clump masses $\log_{10}m_{ij}^{t}$
  ($i=0,1,\cdots,49$ and $j=0,1,\cdots,49$).
 	The horizontal axis represents the logarithm of the expected value $\mu$ to base 2, $\log_{2}\mu$.
 	The values of the initial total mass and the time are as follows: (a)$m_{G}^{0}=0.501$, $t=1570$; (b)$m_{G}^{0}=0.999$, $t=660$; (c)$m_{G}^{0}=2.00$, $t=800$; (d)$m_{G}^{0}=4.01$, $t=90$.
  }
  \label{fig:pd.eps}
\end{figure}

By the fast computation of the proposed CML,
we have performed simulations over a wide range of the parameters and the initial conditions.
The simulation results differ widely depending on the expected value of initial total mass $\mu$, as shown in Fig.\,\ref{fig:pd.eps}.
The expected value $\mu$ is an important parameter to search for possible patterns induced by gravity.
It determines how much material of astronomical objects (that is, gas) exists in a space at $t=0$.
The amount of the material changes the patterns of astronomical objects appearing in the space
since it gives the magnitude of gravitational interaction among gas clumps.
%since its increase leads to a strong gravitational interaction between gas clumps.
From another point of view, we have a conjecture that
the increase of $\mu$ leads to the patterns of astronomical objects which appear on a larger scale ($\propto\sqrt{\mu}$), without our changing the lattice size.
%, under the following assumption.
%We introduce a space where the distance $l$ corresponds to the distance of one between the nearest neighbor lattice points.
%Assuming that the density of the space $\mu/(N_{x}N_{y}l^{2})$ is constant with respect to $\mu$, the size of the space $N_{x}N_{y}l^{2}$ increases proportionally to $\mu$.
In the universe, different patterns are indeed observed with the increase of the scale,
such as galaxies (see Fig.\,\ref{fig:pd.eps}b), galaxy groups (see Fig.\,\ref{fig:pd.eps}d), galaxy clusters and galaxy filaments\cite{GalaDy}.
%Here we note that this assumption seems to be reasonable since it leads to the scale invariance of the standard deviation $\sigma$ ($\sigma\propto l^{2}$)\cite{GalaDy}.

The results are roughly classified into the following four types based on the mass and number of the formed stars:
(1) When $\mu=0.5$ (see Fig.\,\ref{fig:pd.eps}a),
gas clumps form a light central star (the mass of about 0.5),
and it cannot gravitate gas particles strongly,
and they leak out before ejected from it,
and thus an unusual spiral pattern is formed, such as with a pair of asymmetry arms;
(2) When $\mu=1$ (see Fig.\,\ref{fig:pd.eps}b),
gas clumps form a central star (the mass of about 1),
and it ejects gas particles at its maximal contraction,
and thus a grand design spiral pattern is formed;
(3) When $\mu=2$ (see Fig.\,\ref{fig:pd.eps}c),
gas clumps form a massive central star (the mass of about 2),
and it is repeatedly deformed due to gravitational instability
and constantly ejecting gas particles in all directions,
and thus a spiral pattern is not formed;
(4) When $\mu=4$ (see Fig.\,\ref{fig:pd.eps}d),
gas clumps gather to form two or more stars (the mass of about 1 to 3),
and all of them form spiral patterns as (2), or do not form as (3),
or some of them form as (2) and others do not as (3),
and then they change into more complex and diverse patterns due to gravitational interaction among them.

The variance of initial total mass $\sigma^{2}$ (and that of $m_{ij}^{0}$) is also important in the above pattern formation.
When the expected value $\mu$ is small (in the patterns of (1) to (3)),
the variance $\sigma^{2}$ gives the fluctuations of the mass, momentum and angular momentum of a formed single star,
and the fluctuations result in gas ejection from the star,
and thus patterns are formed as described above.
When the expected value $\mu$ is large (in the patterns of (4)),
the variance $\sigma^{2}$ additionally gives the fluctuation even of the number of formed (multiple) stars,
and patterns vary diversely due to the growth of these fluctuations accelerated by the chaotic motion of the stars.

\subsection{Formation process of grand design spiral patterns}
\label{Formation process of grand design spiral patterns}

We report a new formation process of grand design spiral patterns in the above simulation results of (2),
following the snapshots at $t=0,60,\cdots,$ and $1860$ in Fig.\,\ref{fig:alpha_0.eps},
and also at $t=430,440,\cdots,$ and $480$ in Fig.\,\ref{fig:emf.eps}. 
In Sections \ref{Formation of a central star} and \ref{Formation of spiral patterns},
the time in brackets shows the exact time when the focused event occurs or is occurring.

\begin{figure*}[t]
  \begin{center}
    \includegraphics[scale=0.5]{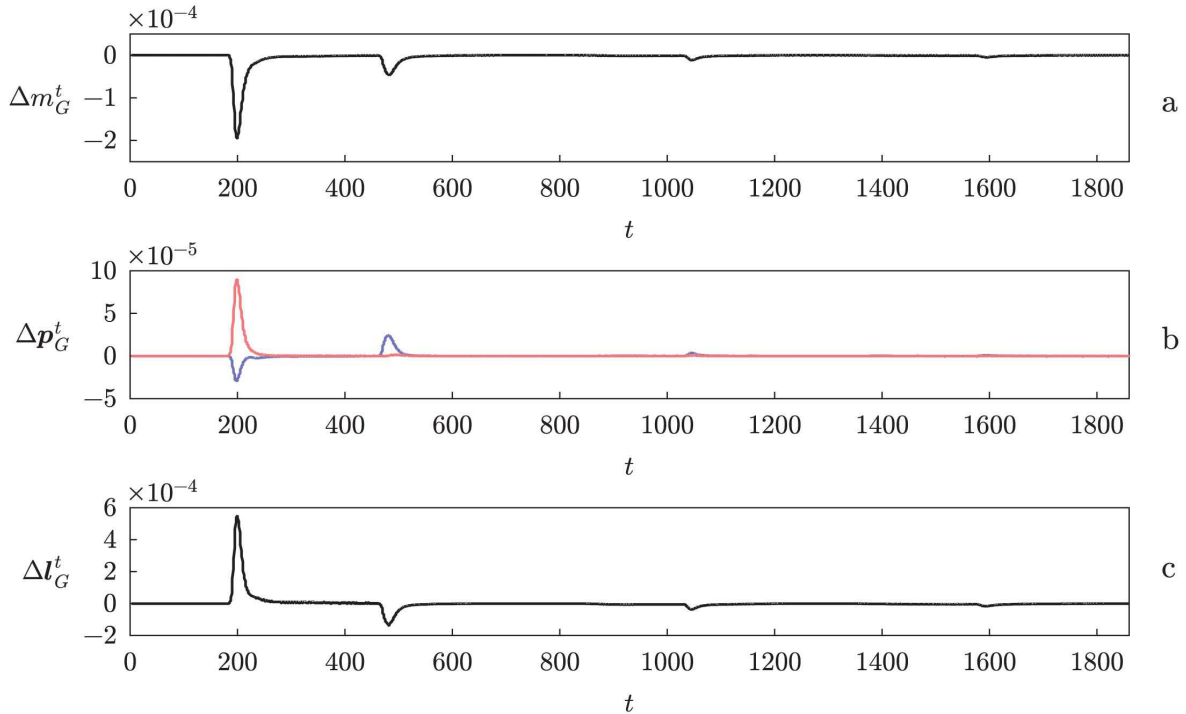}
  \end{center}
  \caption{
	Time series of the time rate of change of the constant quantities of the system.
	(a) Time rate of change of the total mass $\Delta m_{G}^{t}$.
	(b) Time rate of change of the total momentum $\Delta \Vec{p}_{G}^{t}$.
	The blue line represents the $x$-component and the red line the $y$-component.
	(c) Time rate of change of the total angular momentum $\Delta \Vec{l}_{G}^{t}$.
	}
  \label{fig:toma.eps}
\end{figure*}

\subsubsection{Formation of a central star}
\label{Formation of a central star}

A central star is formed from gas clumps of random initial masses, through the following sequence of events.
Gas clumps gravitate to each other and rapidly move to the center of the lattice,
as shown in the snapshots at $t=0,60$ and $120$.
Almost all of the gas clumps converge into a cluster of the four massive gas clumps ($99.2\%$ of the initial total mass)
at lattice points $24\ 24$, $25\ 24$, $24\ 25$ and $25\ 25$ respectively,
as in the snapshot at $t=180$ (at $t=153$ in the exact time).
This is the birth of the central star consisting of four massive gas clumps.

The central star starts to move approximately in the positive $y$-direction and rotate counterclockwise,
through the following sequence of events.
The tiny part of the gas clumps ($0.8\%$ of the initial total mass)
does not fall into the central star,
but swings near it and leaves it, like a spacecraft in a gravitational slingshot.
One part of the leaving gas clumps ($0.4\%$ of the initial total mass)
moves out through the boundary of the finite lattice,
as in the snapshots at $t=180,240$ and $300$ (from $t=179$ to $t=281$ in the exact time).
Here we note that it appears as the white thick line from the central star to the lower boundary.
Thus, in this particular simulation,
the central star (that is, the system) gains a momentum in the positive $y$-direction
and also a little in the negative $x$-direction (see Fig.\,\ref{fig:toma.eps}b),
and an angular momentum in the positive $z$-direction (counterclockwise rotation, see Fig.\,\ref{fig:toma.eps}c),
where the $z$-direction is defined by unit vector $\Vec{e}_{z}=\Vec{e}_{x}\times\Vec{e}_{y}$.
Meanwhile, the other part of the leaving gas clumps ($0.4\%$ of the initial total mass)
gravitates to the central star again, as in the snapshots at $t=360$ and $420$.

\subsubsection{Formation of grand design spiral patterns}
\label{Formation of spiral patterns}

The first grand design spiral pattern is formed originating in gas ejection from the central star, through the following sequence of events.
The four massive gas clumps in the central star ($99.6\%$ of the initial total mass) maximally contract to two massive ones
in the positive $y$-direction by the movement of the central star.
Then each one of the two massive gas clumps ejects gas particles ($0.6\%$ of the central star mass),
as in the snapshots of Fig.\,\ref{fig:emf.eps} at $t=430, 440$ and $450$ (from $t=431$ to $t=448$ in the exact time),
since the central star increases its counterclockwise rotational speed as it contracts
under the conservation of the total angular momentum,
like a spinning figure skater whose arms are contracted.
Most of the ejected gas particles start Keplerian motion around the central star
and the Keplerian gas particles become jammed while passing through the two high density areas of them,
as in the snapshots of Fig.\,\ref{fig:emf.eps} at $t=460, 470$ and $480$.
The jammed Keplerian gas particles form a pair of spiral arms and the first grand design spiral pattern appears,
as in the snapshots at $t=480,540,\cdots,$ and $960$.
The spiral pattern becomes larger and less sharp gradually as time evolves,
while slowly rotating counterclockwise around the central star.

The central star still moves almost in the positive $y$-direction and rotates counterclockwise
after the first gas ejection from the central star, through the following sequence of events.
A small fraction of the ejected gas particles ($0.1\%$ of the initial total mass)
moves out through the boundary of the lattice (from $t=465$ to $t=575$ in the exact time),
and hence the central star now gains a momentum
in the positive $x$-direction (see Fig.\,\ref{fig:toma.eps}b)
and an angular momentum in the negative $z$-direction (clockwise rotation, see Fig.\,\ref{fig:toma.eps}c).
As shown in Fig.\,\ref{fig:toma.eps}b,
the positive $x$-component of the momentum which the central star gains at this time
almost cancels out the previous negative $x$-component
and thus the central star has the total momentum almost only in the positive $y$-direction.
In addition, as shown in Fig.\,\ref{fig:toma.eps}c,
the negative $z$-component of the angular momentum which the central star gains at this time
is smaller than the previous positive $z$-component
and thus the central star still has the total angular momentum in the positive $z$-direction (counterclockwise rotation).

The second grand design spiral pattern is also formed through the following sequence of events.
The central star ($99.3\%$ of the initial total mass) ejects gas particles ($0.2\%$ of the central star mass) again
in the same way as the first spiral pattern,
as in the snapshots at $t=960$ and $1020$ (from $t=990$ to $t=995$ in the exact time).
The ejected gas particles almost do not move out through the boundary (see Fig.\,\ref{fig:toma.eps}a)
since they collide with Keplerian gas particles ($0.2\%$ of the central star mass) of the first spiral pattern.
They jam the Keplerian gas particles
and the jammed Keplerian gas particles around the central star form the second grand design spiral pattern,
as in the snapshots at $t=1020,1080,\cdots,$ and $1500$.
The second spiral pattern shows not only the same behavior as the first spiral pattern
(to rotate around the central star and become larger by degrees)
but also a different behavior of keeping its two arms sharp for a longer time.

The third grand design spiral pattern is also formed through the following sequence of events.
The central star ($99.2\%$ of the initial total mass) ejects gas particles ($0.1\%$ of the central star mass)
once again in the same way,
as in the snapshots at $t=1500$ and $1560$ (from $t=1529$ to $t=1532$ in the exact time).
The ejected gas particles hardly move out through the boundary (see Fig.\,\ref{fig:toma.eps}a)
due to the collision with Keplerian gas particles ($0.3\%$ of the central star mass) of the second spiral pattern.
The jammed Keplerian gas particles around the central star form the third grand design spiral pattern,
as in the snapshots at $t=1560,1620,\cdots,$ and $1860$.
The third spiral pattern is quite similar to the second one,
such as the snapshots at $t=1560$ and $1020$, $1620$ and $1080$, $\cdots$, or $1860$ and $1320$.
It also shows almost the same behavior as the second one
which rotates around the central star, becomes larger gradually and keeps the two sharp arms.

Even after the third formation, the similar formation of spiral patterns is repeated over and over again at intervals of around 540 steps.
However, the formation of spiral patterns finishes in the end
since the central star does not eject gas particles by losing its angular momentum at each time of gas ejection.

In the above formation process,
the spiral patterns are grand design spirals which have two gaseous spiral arms.
They have been universally found in not only grand design but also flocculent spiral galaxies\cite{Kuno}.
Moreover, the Keplerian motion of gas particles is a fundamental dynamic property of astronomical objects\cite{GalaDy} (for details, see the next section),
and in agreement with the observations near the center of the spiral galaxy NGC4258 (M106) \cite{Miyoshi} and those in the protoplanetary disk of VLA1623A\cite{Nadia}.
Furthermore, the formation of spiral arms by jammed Keplerian gas particles is a property of density waves\cite{Lin-Shu},
and in qualitative agreement with the observations in the spiral galaxy M51\cite{Kuno2}.
In addition to this agreement, it has been observed that narrow jets and wide outflows are accelerated by independent mechanisms\cite{Tombesi,Matsushita}.
Thus, it should be suggested that such gas ejection as described above can happen and lead to the formation of grand design spiral patterns, in actual spiral galaxies and protoplanetary disks.

\begin{figure*}[t]
  \begin{center}
    \includegraphics[scale=0.95]{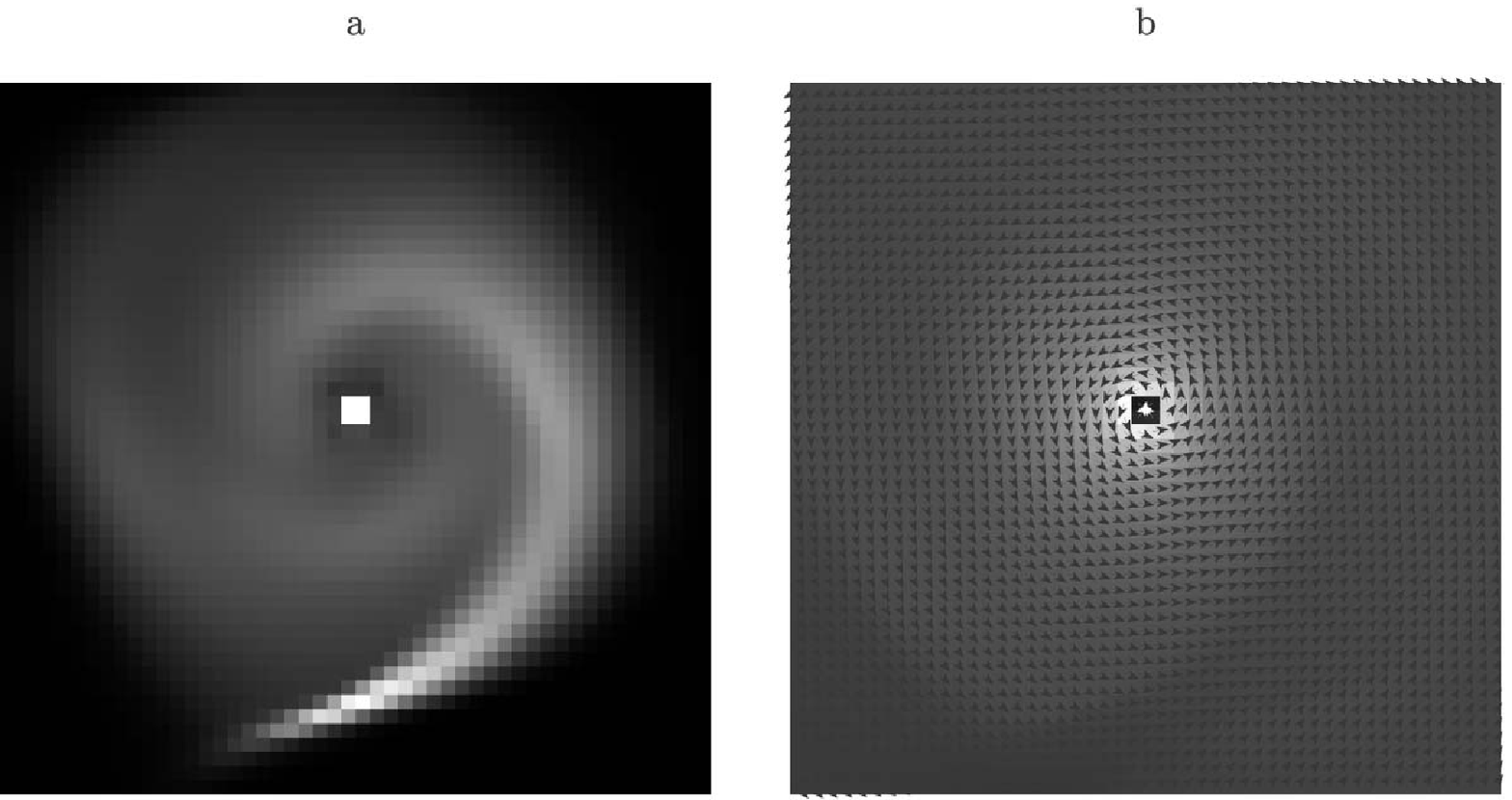}\vspace{10pt}
    \includegraphics[scale=0.475]{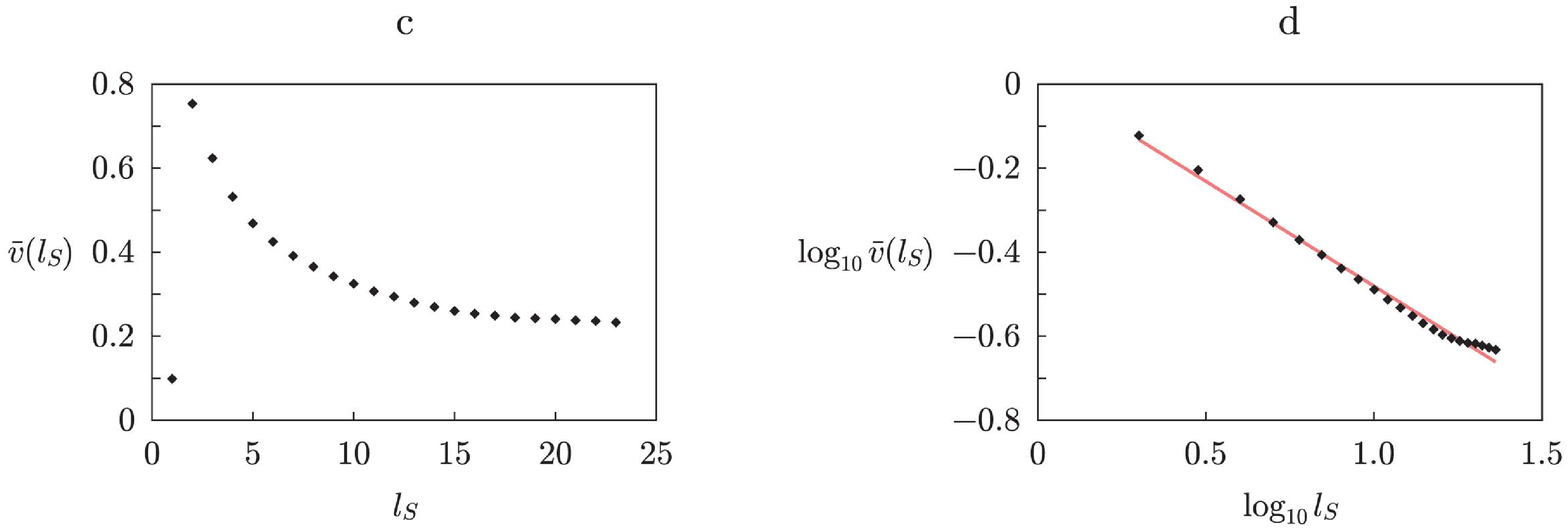}
  \end{center}
  \caption{
	Keplerian motion of gas clumps at $t=1360$.
	(a) Snapshot of gas clump masses $m_{ij}^{t}$. 
	The brightness of cells represents gas clump masses $m_{ij}^{t}$ in the range of 0 to $2\times 10^{-6}$.
	(b) Snapshot of gas clump velocities $\Vec{v}_{ij}^{t}$.
	The brightness of cells represents gas clump speeds $\left|\Vec{v}_{ij}^{t}\right|$ in the range of 0.06 to 0.84
	and the black arrows the direction of gas clump velocities $\Vec{v}_{ij}^{t}$.
	(c) Rotation curve.
	The horizontal axis represents distance $l_{S}$ from $\Vec{r}_{S}^{t}$
	and vertical axis average speed $\bar{v}(l_{S})$.
	(d) Log-log plot of the rotation curve. The red line represents its regression line.
  }
  \label{fig:mfvf.eps}
\end{figure*}

\section{Keplerian motion of gas clumps}
\label{Keplerian motion of gas clumps}

Gas clumps are in Keplerian motion around the central star,
which is one of the most fundamental dynamic properties of astronomical objects.
We verify it by using the snapshot of gas clump velocities $\Vec{v}_{ij}^{t}$ and the rotation curve of a spiral pattern
shown in Fig.\,\ref{fig:mfvf.eps}.

First, the snapshot of gas clump velocities $\Vec{v}_{ij}^{t}$ suggests the Keplerian motion of gas clumps.
When two spiral arms are formed as shown in Fig.\,\ref{fig:mfvf.eps}a,
the black arrows in Fig.\,\ref{fig:mfvf.eps}b represent the direction of gas clump velocities $\Vec{v}_{ij}^{t}$
and the brightness of cells in Fig.\,\ref{fig:mfvf.eps}b the gas clump speeds $\left|\Vec{v}_{ij}^{t}\right|$.
We find that each gas clump is in elliptic motion around the central star, following the black arrows.
We also find that gas clump speeds $\left|\Vec{v}_{ij}^{t}\right|$ decrease with the increase of the distance from the central star,
comparing the brightness of cells.
Thus gas clump velocities $\Vec{v}_{ij}^{t}$ have the properties of Keplerian motion.

Next, we draw the rotation curve of the spiral pattern in order to show it more clearly.
The rotation curve is expressed as
the plot of the average speed $\bar{v}(l_{S})$ of gas clumps at the distance $l_{S}$
from the center of gravity of the central star $\Vec{r}_{S}^{t}$.
The average speed $\bar{v}(l_{S})$ is given by
averaging gas clump speeds $\left|\Vec{v}_{ij}^{t}\right|$ over the gas clumps
whose positions $\Vec{r}_{ij}$ satisfy $l_{S}\le \left|\Vec{r}_{ij}-\Vec{r}_{S}^{t}\right|<l_{S}+1$.

Figs.\,\ref{fig:mfvf.eps}c and \ref{fig:mfvf.eps}d show
the rotation curve of the spiral pattern and its log-log plot, respectively.
In Fig.\,\ref{fig:mfvf.eps}d,
the log-log plot of the rotation curve decreases linearly with distance $l_{S}$
and its regression line (the red line in Fig.\,\ref{fig:mfvf.eps}d) is given by
\begin{equation}
\log_{10} \bar{v}(l_{S})=-5.0\times 10^{-1}\log_{10}l_{S}+1.9\times 10^{-2}.
\end{equation}
The fit between the regression line and the rotation curve is good
and the residual sum of squares is $3.7\times 10^{-3}$.
Thus average speed $\bar{v}(l_{S})$ becomes
\begin{equation}
\label{eq:v(lS)}
\bar{v}(l_{S})\sim \sqrt{\frac{1}{l_{S}}}
\end{equation}
and therefore it is verified that gas clumps are in Keplerian motion around the central star.

It is known that the rotation curves of spiral galaxies do not keep decreasing as Fig.\,\ref{fig:mfvf.eps} away from the center of the galaxies,
due to the existence of dark matter\cite{GalaDy} which is not taken into account in the proposed CML.
The modified CML which considers the effect of dark matter will be reported elsewhere.

\section{Summary and Discussion}
\label{summary}

\begin{figure*}[t]
  \begin{center}
    \includegraphics[scale=0.5]{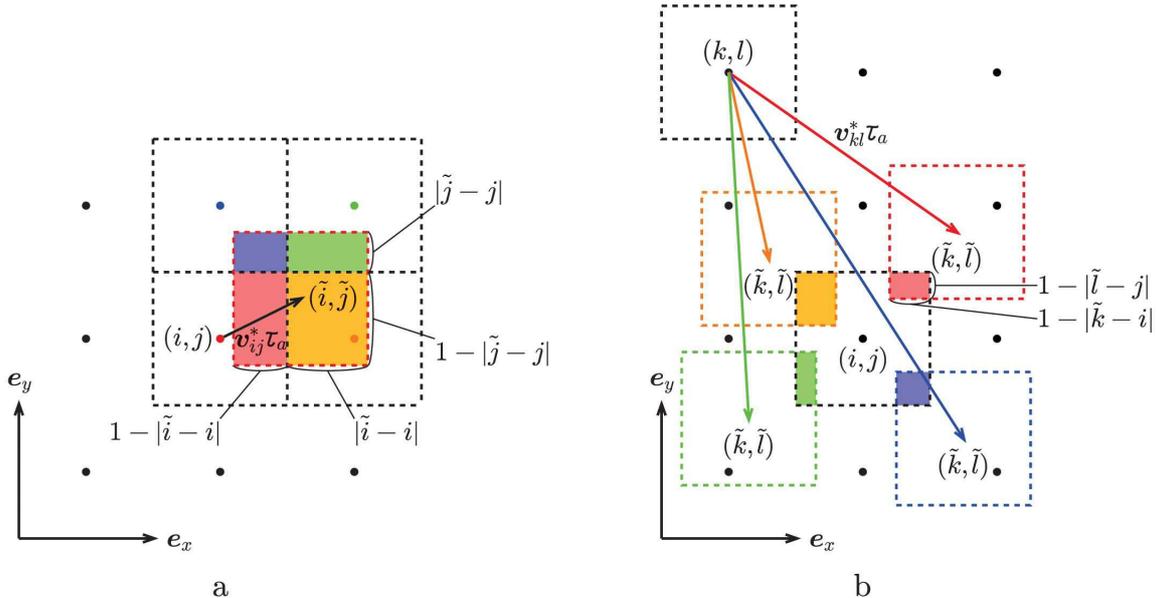}
  \end{center}
  \caption{
	Advection in the particle picture.
  (a) Advection from the cell at lattice point $ij$.
  (b) Advection from the cell at lattice point $kl$ to the cell at lattice point $ij$.
  }
  \label{fig:advection.eps}
\end{figure*}

We have proposed a minimal CML
for the spiral pattern formation in astronomical objects consisting of accreting gas.
The proposed CML consists of a minimal set of procedures:
One is a gravitational interaction procedure as an Eulerian procedure
and the other an advection procedure as a Lagrangian procedure.
Through the simulation, we have found that those two procedures
are a necessary (and therefore essential) set to form grand design spiral patterns,
such as seen in spiral galaxies or protoplanetary disks.

We have observed the following new formation process of grand design spiral patterns in the simulation of the proposed CML.
First, the formed central star starts to move and rotate.
It contracts in the direction of movement and increases its rotational speed
under the conservation of the total angular momentum.
Then it ejects gas particles when it maximally contracts.
The ejected gas particles become a jammed Keplerian gas particles around the central star.
Finally, the jammed gas particles form a grand design spiral pattern.

We have found that
three aspects of the formation process agree well with the results of the conventional theories\cite{GalaDy,Lin-Shu} and observations\cite{Kuno,Miyoshi,Nadia,Kuno2}.
The grand design spiral patterns are universally found in not only grand design but also flocculent spiral galaxies\cite{Kuno}.
The motion of gas particles in the spiral patterns is Keplerian\cite{GalaDy} as shown in the analysis using the rotation curve, and agrees with the observations near the center of the spiral galaxy NGC4258 (M106) \cite{Miyoshi} and those in the protoplanetary disk of VLA1623A\cite{Nadia}.
The formation of spiral arms by jammed Keplerian gas particles is a property of density waves\cite{Lin-Shu}, and also agrees with those in the spiral galaxy M51\cite{Kuno2}.
In addition to this agreement, the observations\cite{Tombesi,Matsushita} have revealed that narrow jets and wide outflows are accelerated by independent mechanisms.
It should be thus suggested that the gas ejection can occur and result in the formation of grand design spiral patterns, in actual spiral galaxies and protoplanetary disks.

We consider that the transition
from one spiral pattern (ordered state) to another via gas ejection (disordered state) from the central star
would be related to chaotic itinerancy\cite{Kaneko,Tsuda} in high-dimensional dynamical systems\cite{Yanagitac,Nozawa,Konishi}.
Moreover,
we expect that the evolution from spiral patterns to less sharp patterns in the transition
would be concerned with the evolution from star-forming spiral galaxies to non-star-forming elliptical galaxies,
which is called galaxy "quenching"\cite{Faber,Koyama}.
These will be reported elsewhere.

\section*{Acknowledgement}

The author would like to thank T. Deguchi for valuable comments and critical reading of the manuscript.
I also would like to thank M. Morikawa for extensive discussions about what the proposed CML should be in astrophysics.
I am grateful to T. Tashiro and T. Ootsuka for useful advice from the point of view of astrophysics.
I also would like to offer my special thanks to H. Nozawa for introduction to CML.

\appendix

\section{Advection procedure}
\label{Advection procedure}

Let us now formulate the advection procedure $T_{a}$ as the Lagrangian procedure\cite{Kaneko,Yanagitac} in the particle picture.

Fig.\,\ref{fig:advection.eps}a shows an example of the advection procedure.
In the cell at lattice point $ij$ (the red dot in Fig.\,\ref{fig:advection.eps}a),
gas particles are distributed uniformly and have the same velocity $\Vec{v}_{ij}^{*}$,
such as shown in Fig.\,\ref{fig:fv.eps}a.
After time interval $\tau_{a}$,
gas particles in the cell at lattice point $ij$ move
by the displacement $\Vec{v}_{ij}^{*}\tau_{a}$ (the black arrow in Fig.\,\ref{fig:advection.eps}a)
to the cell-sized area (the red dotted square in Fig.\,\ref{fig:advection.eps}a)
centered at position $(\tilde{i},\tilde{j})=(i+v_{x\, ij}^{*}\tau_{a},j+v_{y\, ij}^{*}\tau_{a})$.

The cell-sized area overlaps four cells at lattice points $ij$, $i+1j$, $i+1j+1$ and $ij+1$
(the red, orange, green and blue dots in Fig.\,\ref{fig:advection.eps}a, respectively).
The sizes of its four overlapping areas (the red, orange, green and blue rectangles in Fig.\,\ref{fig:advection.eps}a, respectively)
are $(1-|\tilde{i}-i|)(1-|\tilde{j}-j|)$, $|\tilde{i}-i|(1-|\tilde{j}-j|)$, $|\tilde{i}-i||\tilde{j}-j|$
and $(1-|\tilde{i}-i|)|\tilde{j}-j|$, respectively.
Gas particles in the cell-sized area are carried to each overlapped cell in proportion to the size of the overlapping area,
and the mass and momentum of the gas particles are also carried there (the allocation by the lever rule\cite{Yanagitac}).

We now consider advection from lattice point $kl$ to $ij$, as shown in Fig.\,\ref{fig:advection.eps}b,
in order to describe the advection procedure $T_{a}$
as the map which acts on gas clump mass $m_{ij}^{*}$ and velocity $\Vec{v}_{ij}^{*}$.
Gas particles in the cell at lattice point $kl$ move by displacement $\Vec{v}_{kl}^{*}\tau_{a}$
to the cell-sized area centered at position $(\tilde{k},\tilde{l})$, which is given by Eq.\,\ref{eqn:kl}.
There are the following four cases of position $(\tilde{k},\tilde{l})$, as shown in Fig.\,\ref{fig:advection.eps}b.
When position $(\tilde{k},\tilde{l})$ is located in the upper right side of lattice point $ij$,
the cell-sized area (the red dotted square in Fig.\,\ref{fig:advection.eps}b) overlaps the cell at lattice point $ij$.
The size of its overlapping area (the red rectangle in Fig.\,\ref{fig:advection.eps}b) is $(1-|\tilde{k}-i|)(1-|\tilde{l}-j|)$.
Gas particles in the cell-sized area are carried to the cell at lattice point $ij$,
with their mass $(1-|\tilde{k}-i|)(1-|\tilde{l}-j|)m_{kl}^{*}$
and momentum $(1-|\tilde{k}-i|)(1-|\tilde{l}-j|)m_{kl}^{*}\Vec{v}_{kl}^{*}$.
In a similar way, when position $(\tilde{k},\tilde{l})$ is located
in the upper left, lower left or lower right side of lattice point $ij$,
the corresponding cell-sized area (the orange dotted square, green or blue in Fig.\,\ref{fig:advection.eps}b, respectively)
overlaps the cell at lattice point $ij$ respectively.
The size of its overlapping area (the orange rectangle, green or blue in Fig.\,\ref{fig:advection.eps}b, respectively)
is $(1-|\tilde{k}-i|)(1-|\tilde{l}-j|)$.

Thus, the weight of allocation $w_{ijkl}^{*}$ of gas particles from lattice point $kl$ to $ij$
is given as Eq.\,\ref{eqn:w_ijkl}.
In Eq.\,\ref{eqn:w_ijkl},
the first term in the first brackets gives the flag to become one
when position $(\tilde{k},\tilde{l})$ is located in the upper right of lattice point $ij$,
the second term the upper left, the third term the lower left and the fourth term the lower right.

After carried from different cells at lattice points $kl$ to the cell at lattice point $ij$, as described above,
gas particles collide with each other in the cell at lattice point $ij$ and form a new gas clump
(they are distributed uniformly and have the same velocity, as shown in Fig.\ref{fig:fv.eps}a).
In this collision process, the mass, momentum and angular momentum of the carried gas particles are conserved.

Through the above discussion, we introduce the advection procedure $T_{a}$ as Eqs.\,\ref{eqn:Ta_m} and \ref{eqn:Ta_v}.

\section{Computational cost of the advection procedure}
\label{Computational cost of advection procedure}

In the operation of the advection procedure of Eqs.\,\ref{eqn:Ta_m} and \ref{eqn:Ta_v},
we can easily show that the computational cost is given by only $O(N)$,
where $N$ is the total number of lattice points.
Eq.\,\ref{eqn:Ta_m} is rewritten as
\begin{equation}
\label{eqn:mt+1}
\Vec{m}^{t+1}=W^{*}\Vec{m}^{t}
\end{equation}
with
\begin{eqnarray}
\Vec{m}^{t}=m_{00}^{t}\Vec{e}_{00}+\cdots+m_{ij}^{t}\Vec{e}_{ij}+\cdots
\nonumber\\
\lefteqn{+m_{N_{x}-1N_{y}-1}^{t}\Vec{e}_{N_{x}-1N_{y}-1},}
\hphantom{\Vec{m}^{t}=m_{00}^{t}\Vec{e}_{00}+\cdots+m_{ij}^{t}\Vec{e}_{ij}+\cdots}
\end{eqnarray}
\begin{eqnarray}
W^{*}=
\left(
\begin{array}{ccc}
w_{00\, 00}^{*} & \cdots & w_{00\, kl}^{*} \\
\vdots & & \vdots \\
w_{ij\, 00}^{*} & \cdots & w_{ij\, kl}^{*} \\
\vdots & & \vdots \\
w_{N_{x}-1N_{y}-1\, 00}^{*} & \cdots & w_{N_{x}-1N_{y}-1\, kl}^{*} \\
\end{array}
\right.
\nonumber\\
\lefteqn{\left.
\begin{array}{cc}
\cdots & w_{00\, N_{x}-1N_{y}-1}^{*} \\
 & \vdots \\
\cdots & w_{ij\, N_{x}-1N_{y}-1}^{*} \\
 & \vdots \\
\cdots & w_{N_{x}-1N_{y}-1\, N_{x}-1N_{y}-1}^{*} \\
\end{array}
\right),}
\hphantom{W^{*}=
\left(
\begin{array}{ccc}
w_{00\, 00}^{*} & \cdots & w_{00\, kl}^{*} \\
\vdots & & \vdots \\
w_{ij\, 00}^{*} & \cdots & w_{ij\, kl}^{*} \\
\vdots & & \vdots \\
w_{N_{x}-1N_{y}-1\, 00}^{*} & \cdots & w_{N_{x}-1N_{y}-1\, kl}^{*} \\
\end{array}
\right.
}
\end{eqnarray}
where $\Vec{e}_{ij}$ is a standard unit vector.
Eq.\,\ref{eqn:mt+1} suggests a high computational cost of $O(N^{2})$ in the operation of the advection procedure.
However, it is not the case.
Note that, from Eq.\,\ref{eqn:w_ijkl}, matrix $W^{*}$ is a sparse matrix
which has only four (at most) non-zero values in each column
\begin{eqnarray}
w_{\lfloor \tilde{k}\rfloor\lfloor \tilde{l}\rfloor kl}^{*}
=\left\{1-\left(\tilde{k}-\lfloor \tilde{k}\rfloor\right)\right\}
\left\{1-\left(\tilde{l}-\lfloor \tilde{l}\rfloor\right)\right\},\\
\lefteqn{w_{\lfloor \tilde{k}\rfloor +1\lfloor \tilde{l}\rfloor kl}^{*}
=\left(\tilde{k}-\lfloor \tilde{k}\rfloor\right)\left\{1-\left(\tilde{l}-\lfloor \tilde{l}\rfloor\right)\right\},}
\hphantom{w_{\lfloor \tilde{k}\rfloor\lfloor \tilde{l}\rfloor kl}^{*}
=\left\{1-\left(\tilde{k}-\lfloor \tilde{k}\rfloor\right)\right\}
\left\{1-\left(\tilde{l}-\lfloor \tilde{l}\rfloor\right)\right\},}\\
\lefteqn{w_{\lfloor \tilde{k}\rfloor +1\lfloor \tilde{l}\rfloor +1 kl}^{*}
=\left(\tilde{k}-\lfloor \tilde{k}\rfloor\right)\left(\tilde{l}-\lfloor \tilde{l}\rfloor\right),}
\hphantom{w_{\lfloor \tilde{k}\rfloor\lfloor \tilde{l}\rfloor kl}^{*}
=\left\{1-\left(\tilde{k}-\lfloor \tilde{k}\rfloor\right)\right\}
\left\{1-\left(\tilde{l}-\lfloor \tilde{l}\rfloor\right)\right\},}\\
\lefteqn{w_{\lfloor \tilde{k}\rfloor\lfloor \tilde{l}\rfloor +1 kl}^{*}
=\left\{1-\left(\tilde{k}-\lfloor \tilde{k}\rfloor\right)\right\}\left(\tilde{l}-\lfloor \tilde{l}\rfloor\right).}
\hphantom{w_{\lfloor \tilde{k}\rfloor\lfloor \tilde{l}\rfloor kl}^{*}
=\left\{1-\left(\tilde{k}-\lfloor \tilde{k}\rfloor\right)\right\}
\left\{1-\left(\tilde{l}-\lfloor \tilde{l}\rfloor\right)\right\},}
\end{eqnarray}
We thus have
\begin{eqnarray}
\label{eqn:w*mt}
\lefteqn{W^{*}\Vec{m}^{t}
=W^{*}\sum_{kl}m_{kl}^{t}\Vec{e}_{kl}
=\sum_{kl}m_{kl}^{t}W^{*}\Vec{e}_{kl}}
\hphantom{\left.
+w_{\lfloor \tilde{k}\rfloor +1\lfloor \tilde{l}\rfloor +1 kl}^{*}\Vec{e}_{\lfloor \tilde{k}\rfloor +1\lfloor \tilde{l}\rfloor +1}
+w_{\lfloor \tilde{k}\rfloor\lfloor \tilde{l}\rfloor +1 kl}^{*}\Vec{e}_{\lfloor \tilde{k}\rfloor\lfloor \tilde{l}\rfloor +1}
\right).\hspace{-8.4pt}}
\nonumber\\
\lefteqn{=\sum_{kl}m_{kl}^{t}\left(
w_{\lfloor \tilde{k}\rfloor\lfloor \tilde{l}\rfloor kl}^{*}\Vec{e}_{\lfloor \tilde{k}\rfloor\lfloor \tilde{l}\rfloor}
+w_{\lfloor \tilde{k}\rfloor +1\lfloor \tilde{l}\rfloor kl}^{*}\Vec{e}_{\lfloor \tilde{k}\rfloor +1\lfloor \tilde{l}\rfloor}
\right.}
\hphantom{\left.
+w_{\lfloor \tilde{k}\rfloor +1\lfloor \tilde{l}\rfloor +1 kl}^{*}\Vec{e}_{\lfloor \tilde{k}\rfloor +1\lfloor \tilde{l}\rfloor +1}
+w_{\lfloor \tilde{k}\rfloor\lfloor \tilde{l}\rfloor +1 kl}^{*}\Vec{e}_{\lfloor \tilde{k}\rfloor\lfloor \tilde{l}\rfloor +1}
\right).\hspace{-8.4pt}}
\nonumber\\
\lefteqn{\left.
+w_{\lfloor \tilde{k}\rfloor +1\lfloor \tilde{l}\rfloor +1 kl}^{*}\Vec{e}_{\lfloor \tilde{k}\rfloor +1\lfloor \tilde{l}\rfloor +1}
+w_{\lfloor \tilde{k}\rfloor\lfloor \tilde{l}\rfloor +1 kl}^{*}\Vec{e}_{\lfloor \tilde{k}\rfloor\lfloor \tilde{l}\rfloor +1}
\right).}
\hphantom{\left.
+w_{\lfloor \tilde{k}\rfloor +1\lfloor \tilde{l}\rfloor +1 kl}^{*}\Vec{e}_{\lfloor \tilde{k}\rfloor +1\lfloor \tilde{l}\rfloor +1}
+w_{\lfloor \tilde{k}\rfloor\lfloor \tilde{l}\rfloor +1 kl}^{*}\Vec{e}_{\lfloor \tilde{k}\rfloor\lfloor \tilde{l}\rfloor +1}
\right).\hspace{-8.4pt}}
\nonumber\\
\end{eqnarray}
Eq.\,\ref{eqn:w*mt} takes a low computational cost of only $O(N)$.
The computational cost of Eq.\,\ref{eqn:Ta_v} also can be reduced to $O(N)$ in a similar way.

\section{Conservation of total mass, momentum and angular momentum}
\label{Conservative quantities}

Here we show that the total mass $m_{G}^{t}$, total momentum $\Vec{p}_{G}^{t}$
and total angular momentum $\Vec{l}_{G}^{t}$ of the system are conserved on the time evolution.

First, we verify that these quantities are conserved in the gravitational interaction procedure $T_{g}$.
From Eq.\,\ref{eqn:Tg_m}, total mass $m_{G}^{t}$ becomes
\begin{equation}
m_{G}^{*}
=\sum_{i=0}^{N_{\mathstrut x}-1}\sum_{j=0}^{N_{\mathstrut y}-1}m_{ij}^{*}
=\sum_{i=0}^{N_{\mathstrut x}-1}\sum_{j=0}^{N_{\mathstrut y}-1}m_{ij}^{t}
=m_{G}^{t}.
\end{equation}
Therefore it does not change in the procedure $T_{g}$.
From Eqs.\,\ref{eqn:Tg_m} and \ref{eqn:Tg_v}, total momentum $\Vec{p}_{G}^{*}$ becomes
\begin{eqnarray}
\lefteqn{\Vec{p}_{G}^{*}
=\sum_{i=0}^{N_{\mathstrut x}-1}\sum_{j=0}^{N_{\mathstrut y}-1}m_{ij}^{*}\Vec{v}_{ij}^{*}}
\hphantom{=\sum_{i=0}^{N_{\mathstrut x}-1}\sum_{j=0}^{N_{\mathstrut y}-1}
\left(m_{ij}^{*}\Vec{v}_{ij}^{*}
+\tau_{g}\sum_{k=0}^{N_{\mathstrut x}-1}\sum_{l=0}^{N_{\mathstrut y}-1}\Vec{f}_{ijkl}^{t}\right)}\nonumber\\
=\sum_{i=0}^{N_{\mathstrut x}-1}\sum_{j=0}^{N_{\mathstrut y}-1}
\left(m_{ij}^{t}\Vec{v}_{ij}^{t}
+\tau_{g}\sum_{k=0}^{N_{\mathstrut x}-1}\sum_{l=0}^{N_{\mathstrut y}-1}\Vec{f}_{ijkl}^{t}\right)\nonumber\\
\lefteqn{=\sum_{i=0}^{N_{\mathstrut x}-1}\sum_{j=0}^{N_{\mathstrut y}-1}m_{ij}^{t}\Vec{v}_{ij}^{t}=\Vec{p}_{G}^{t}.}
\hphantom{=\sum_{i=0}^{N_{\mathstrut x}-1}\sum_{j=0}^{N_{\mathstrut y}-1}
\left(m_{ij}^{*}\Vec{v}_{ij}^{*}
+\tau_{g}\sum_{k=0}^{N_{\mathstrut x}-1}\sum_{l=0}^{N_{\mathstrut y}-1}\Vec{f}_{ijkl}^{t}\right)}
\end{eqnarray}
Thus it does not change in the procedure $T_{g}$, either.
Here $\Vec{f}_{ijkl}^{t}$ is
the force of gravity exerted on the gas clump at lattice point $ij$ by the gas clump at lattice point $kl$, given by
\begin{equation}
\Vec{f}_{ijkl}^{t}
=-\gamma(1-\delta_{ik}\delta_{jl})\frac{m_{ij}^{t}m_{kl}^{t}}{|\Vec{r}_{ij}-\Vec{r}_{kl}|^{2}}
\frac{\Vec{r}_{ij}-\Vec{r}_{kl}}{|\Vec{r}_{ij}-\Vec{r}_{kl}|},
\end{equation}
and satisfies
\begin{equation}
\sum_{i,j}\sum_{k,l}\Vec{f}_{ijkl}^{t}
=\frac{1}{2}\sum_{i,j}\sum_{k,l}\left(\Vec{f}_{ijkl}^{t}+\Vec{f}_{klij}^{t}\right)
=\Vec{0}.
\end{equation}
From Eqs.\,\ref{eqn:Tg_m} and \ref{eqn:Tg_v}, total angular momentum $\Vec{l}_{G}^{*}$ becomes
\begin{eqnarray}
\lefteqn{\Vec{l}_{G}^{*}
=\sum_{i=0}^{N_{\mathstrut x}-1}\sum_{j=0}^{N_{\mathstrut y}-1}
\left(\Vec{r}_{ij}-\Vec{r}_{G}^{*}\right)\times m_{ij}^{*}\Vec{v}_{ij}^{*}}
\hphantom{+\tau_{g}\sum_{i=0}^{N_{\mathstrut x}-1}\sum_{j=0}^{N_{\mathstrut y}-1}
\sum_{k=0}^{N_{\mathstrut x}-1}\sum_{l=0}^{N_{\mathstrut y}-1}
\left(\Vec{r}_{ij}-\Vec{r}_{G}^{t}\right)\times \Vec{f}_{ijkl}^{t}}\nonumber\\
\lefteqn{=\sum_{i=0}^{N_{\mathstrut x}-1}\sum_{j=0}^{N_{\mathstrut y}-1}\left(\Vec{r}_{ij}-\Vec{r}_{G}^{t}\right)}
\hphantom{+\tau_{g}\sum_{i=0}^{N_{\mathstrut x}-1}\sum_{j=0}^{N_{\mathstrut y}-1}
\sum_{k=0}^{N_{\mathstrut x}-1}\sum_{l=0}^{N_{\mathstrut y}-1}
\left(\Vec{r}_{ij}-\Vec{r}_{G}^{t}\right)\times \Vec{f}_{ijkl}^{t}}\nonumber\\
\lefteqn{\times\left(m_{ij}^{t}\Vec{v}_{ij}^{t}
+\tau_{g}\sum_{k=0}^{N_{\mathstrut x}-1}\sum_{l=0}^{N_{\mathstrut y}-1}\Vec{f}_{ijkl}^{t}\right)}
\hphantom{+\tau_{g}\sum_{i=0}^{N_{\mathstrut x}-1}\sum_{j=0}^{N_{\mathstrut y}-1}
\sum_{k=0}^{N_{\mathstrut x}-1}\sum_{l=0}^{N_{\mathstrut y}-1}
\left(\Vec{r}_{ij}-\Vec{r}_{G}^{t}\right)\times \Vec{f}_{ijkl}^{t}}\nonumber\\
\lefteqn{=\sum_{i=0}^{N_{\mathstrut x}-1}\sum_{j=0}^{N_{\mathstrut y}-1}
\left(\Vec{r}_{ij}-\Vec{r}_{G}^{t}\right)\times m_{ij}^{t}\Vec{v}_{ij}^{t}}
\hphantom{+\tau_{g}\sum_{i=0}^{N_{\mathstrut x}-1}\sum_{j=0}^{N_{\mathstrut y}-1}
\sum_{k=0}^{N_{\mathstrut x}-1}\sum_{l=0}^{N_{\mathstrut y}-1}
\left(\Vec{r}_{ij}-\Vec{r}_{G}^{t}\right)\times \Vec{f}_{ijkl}^{t}}\nonumber\\
+\tau_{g}\sum_{i=0}^{N_{\mathstrut x}-1}\sum_{j=0}^{N_{\mathstrut y}-1}\sum_{k=0}^{N_{\mathstrut x}-1}\sum_{l=0}^{N_{\mathstrut y}-1}
\left(\Vec{r}_{ij}-\Vec{r}_{G}^{t}\right)\times \Vec{f}_{ijkl}^{t}\nonumber\\
\lefteqn{=\sum_{i=0}^{N_{\mathstrut x}-1}\sum_{j=0}^{N_{\mathstrut y}-1}
\left(\Vec{r}_{ij}-\Vec{r}_{G}^{t}\right)\times m_{ij}^{t}\Vec{v}_{ij}^{t}=\Vec{l}_{G}^{t}.}
\hphantom{+\tau_{g}\sum_{i=0}^{N_{\mathstrut x}-1}\sum_{j=0}^{N_{\mathstrut y}-1}
\sum_{k=0}^{N_{\mathstrut x}-1}\sum_{l=0}^{N_{\mathstrut y}-1}
\left(\Vec{r}_{ij}-\Vec{r}_{G}^{t}\right)\times \Vec{f}_{ijkl}^{t}}
\end{eqnarray}
Therefore it does not change in the procedure $T_{g}$, either.
Here $\Vec{f}_{ijkl}^{t}$ is a central force parallel to $\Vec{r}_{ij}-\Vec{r}_{kl}$
and satisfies
\begin{eqnarray}
\lefteqn{\sum_{i,j}\sum_{k,l}\left(\Vec{r}_{ij}-\Vec{r}_{G}^{t}\right)\times \Vec{f}_{ijkl}^{t}}
\hphantom{=\frac{1}{2}\sum_{i,j}\sum_{k,l}\left(\Vec{r}_{ij}-\Vec{r}_{kl}\right)\times \Vec{f}_{ijkl}^{t}=\Vec{0}}\nonumber\\
\lefteqn{=\frac{1}{2}\sum_{i,j}\sum_{k,l}
\left\{\left(\Vec{r}_{ij}-\Vec{r}_{G}^{t}\right)\times \Vec{f}_{ijkl}^{t}\right.}
\hphantom{=\frac{1}{2}\sum_{i,j}\sum_{k,l}\left(\Vec{r}_{ij}-\Vec{r}_{kl}\right)\times \Vec{f}_{ijkl}^{t}=\Vec{0}}\nonumber\\
\lefteqn{\left.+\left(\Vec{r}_{kl}-\Vec{r}_{G}^{t}\right)\times \Vec{f}_{klij}^{t}\right\}}
\hphantom{=\frac{1}{2}\sum_{i,j}\sum_{k,l}\left(\Vec{r}_{ij}-\Vec{r}_{kl}\right)\times \Vec{f}_{ijkl}^{t}=\Vec{0}}\nonumber\\
=\frac{1}{2}
\sum_{i,j}\sum_{k,l}\left(\Vec{r}_{ij}-\Vec{r}_{kl}\right)\times \Vec{f}_{ijkl}^{t}
=\Vec{0}.
\end{eqnarray}
Thus total mass $m_{G}^{t}$, momentum $\Vec{p}_{G}^{t}$ and angular momentum $\Vec{l}_{G}^{t}$ of the system
are conserved in the gravitational interaction procedure $T_{g}$.

Next, we verify that these quantities are also conserved in the advection procedure $T_{a}$.
Here we focus on not the whole system but the cell at lattice point $ij$
and show that the total mass $m_{ij}^{*}$, total momentum $\Vec{p}_{ij}^{*}=m_{ij}^{*}\Vec{v}_{ij}^{*}$
and total angular momentum $\Vec{l}_{ij}^{*}=\left(\Vec{r}_{ij}-\Vec{r}_{G}^{*}\right)\times m_{ij}^{*}\Vec{v}_{ij}^{*}$
of gas particles in the cell do not change in the procedure $T_{a}$.
The total mass $m_{ij}^{*}$ is allocated to the four nearest neighbor lattice points
$\lfloor \tilde{i}\rfloor\lfloor \tilde{j}\rfloor$, $\lfloor \tilde{i}\rfloor +1\lfloor \tilde{j}\rfloor$,
$\lfloor \tilde{i}\rfloor +1\lfloor \tilde{j}\rfloor +1$ and $\lfloor \tilde{i}\rfloor\lfloor \tilde{j}\rfloor +1$
of the position $(\tilde{i},\tilde{j})=(i+v_{x\, ij}^{*}\tau_{a},j+v_{y\, ij}^{*}\tau_{a})$.
From Eq.\,\ref{eqn:w_ijkl}, the weight of allocation is given by
\begin{eqnarray}
\lefteqn{w_{\lfloor \tilde{i}\rfloor\lfloor \tilde{j}\rfloor ij}^{*}
=\left\{1-\left(\tilde{i}-\lfloor \tilde{i}\rfloor\right)\right\}
\left\{1-\left(\tilde{j}-\lfloor \tilde{j}\rfloor\right)\right\},}
\hphantom{w_{\lfloor \tilde{i}\rfloor\lfloor \tilde{j}\rfloor ij}^{*}
=\left\{1-\left(\tilde{i}-\lfloor \tilde{i}\rfloor\right)\right\}
\left\{1-\left(\tilde{j}-\lfloor \tilde{j}\rfloor\right)\right\},}\\
\lefteqn{w_{\lfloor \tilde{i}\rfloor +1\lfloor \tilde{j}\rfloor ij}^{*}
=\left(\tilde{i}-\lfloor \tilde{i}\rfloor\right)\left\{1-\left(\tilde{j}-\lfloor \tilde{j}\rfloor\right)\right\},}
\hphantom{w_{\lfloor \tilde{i}\rfloor\lfloor \tilde{j}\rfloor ij}^{*}
=\left\{1-\left(\tilde{i}-\lfloor \tilde{i}\rfloor\right)\right\}
\left\{1-\left(\tilde{j}-\lfloor \tilde{j}\rfloor\right)\right\},}\\
\lefteqn{w_{\lfloor \tilde{i}\rfloor +1\lfloor \tilde{j}\rfloor +1 ij}^{*}
=\left(\tilde{i}-\lfloor \tilde{i}\rfloor\right)\left(\tilde{j}-\lfloor \tilde{j}\rfloor\right),}
\hphantom{w_{\lfloor \tilde{i}\rfloor\lfloor \tilde{j}\rfloor ij}^{*}
=\left\{1-\left(\tilde{i}-\lfloor \tilde{i}\rfloor\right)\right\}
\left\{1-\left(\tilde{j}-\lfloor \tilde{j}\rfloor\right)\right\},}\\
\lefteqn{w_{\lfloor \tilde{i}\rfloor\lfloor \tilde{j}\rfloor +1 ij}^{*}
=\left\{1-\left(\tilde{i}-\lfloor \tilde{i}\rfloor\right)\right\}\left(\tilde{j}-\lfloor \tilde{j}\rfloor\right).}
\hphantom{w_{\lfloor \tilde{i}\rfloor\lfloor \tilde{j}\rfloor ij}^{*}
=\left\{1-\left(\tilde{i}-\lfloor \tilde{i}\rfloor\right)\right\}
\left\{1-\left(\tilde{j}-\lfloor \tilde{j}\rfloor\right)\right\},}
\end{eqnarray}
Thus the total mass of the allocated gas particles becomes
\begin{eqnarray}
\sum_{k=\lfloor\tilde{i}\rfloor}^{\lfloor\tilde{i}\rfloor+1}\sum_{l=\lfloor\tilde{j}\rfloor}^{\lfloor\tilde{j}\rfloor+1}
w_{klij}^{*}m_{ij}^{*}
=\left(\sum_{k=\lfloor\tilde{i}\rfloor}^{\lfloor\tilde{i}\rfloor+1}\sum_{l=\lfloor\tilde{j}\rfloor}^{\lfloor\tilde{j}\rfloor+1}
w_{klij}^{*}\right)m_{ij}^{*}\nonumber\\
\lefteqn{=1\cdot m_{ij}^{*}=m_{ij}^{*},}
\hphantom{\sum_{k=\lfloor\tilde{i}\rfloor}^{\lfloor\tilde{i}\rfloor+1}\sum_{l=\lfloor\tilde{j}\rfloor}^{\lfloor\tilde{j}\rfloor+1}
w_{klij}^{*}m_{ij}^{*}
=\left(\sum_{k=\lfloor\tilde{i}\rfloor}^{\lfloor\tilde{i}\rfloor+1}\sum_{l=\lfloor\tilde{j}\rfloor}^{\lfloor\tilde{j}\rfloor+1}
w_{klij}^{*}\right)m_{ij}^{*}}
\end{eqnarray}
and equals the total mass before allocation.
The total momentum $\Vec{p}_{ij}^{*}$ is also allocated
to the four nearest neighbors of position $(\tilde{i},\tilde{j})$.
Thus the total momentum of the allocated gas particles becomes
\begin{equation}
\sum_{k=\lfloor\tilde{i}\rfloor}^{\lfloor\tilde{i}\rfloor+1}\sum_{l=\lfloor\tilde{j}\rfloor}^{\lfloor\tilde{j}\rfloor+1}
w_{klij}^{*}m_{ij}^{*}\Vec{v}_{ij}^{*}
=m_{ij}^{*}\Vec{v}_{ij}^{*}
=\Vec{p}_{ij}^{*},
\end{equation}
and also equals the total momentum before allocation.
We consider the conservation of total angular momentum $\Vec{l}_{ij}^{*}$ after both the movement and allocation of gas particles.
After gas particles in the cell at lattice point $ij$ moves by displacement $\Vec{v}_{ij}^{*}\tau_{a}$,
the total angular momentum of them becomes
\begin{eqnarray}
\lefteqn{\left(\Vec{r}_{ij}+\Vec{v}_{ij}^{*}\tau_{a}-\Vec{r}_{G}^{*}\right)\times m_{ij}^{*}\Vec{v}_{ij}^{*}}
\phantom{=\left(\Vec{r}_{ij}-\Vec{r}_{G}^{*}\right)\times m_{ij}^{*}\Vec{v}_{ij}^{*}=\Vec{l}_{ij}^{*},}\nonumber\\
=\left(\Vec{r}_{ij}-\Vec{r}_{G}^{*}\right)\times m_{ij}^{*}\Vec{v}_{ij}^{*}=\Vec{l}_{ij}^{*},
\end{eqnarray}
and equals the total angular momentum before movement.
After that, these gas particles are allocated to the four nearest neighbors of position $(\tilde{i},\tilde{j})$.
The total angular momentum of them becomes
\begin{eqnarray}
\lefteqn{\sum_{k=\lfloor\tilde{i}\rfloor}^{\lfloor\tilde{i}\rfloor+1}\sum_{l=\lfloor\tilde{j}\rfloor}^{\lfloor\tilde{j}\rfloor+1}
\left(\Vec{r}_{kl}-\Vec{r}_{G}^{*}\right)\times w_{klij}^{*}m_{ij}^{*}\Vec{v}_{ij}^{*}}
\hphantom{+\left(\tilde{i}-\lfloor\tilde{i}\rfloor\right)\Vec{e}_{x}\times m_{ij}^{*}\Vec{v}_{ij}^{*}
+\left(\tilde{j}-\lfloor\tilde{j}\rfloor\right)\Vec{e}_{y}\times m_{ij}^{*}\Vec{v}_{ij}^{*}}\nonumber\\
\lefteqn{=\left(\lfloor\tilde{i}\rfloor\Vec{e}_{x}+\lfloor\tilde{j}\rfloor\Vec{e}_{y}\right)
\times\left(
w_{\lfloor \tilde{i}\rfloor\lfloor \tilde{j}\rfloor ij}^{*}+w_{\lfloor \tilde{i}\rfloor +1\lfloor \tilde{j}\rfloor ij}^{*}\right.}
\hphantom{+\left(\tilde{i}-\lfloor\tilde{i}\rfloor\right)\Vec{e}_{x}\times m_{ij}^{*}\Vec{v}_{ij}^{*}
+\left(\tilde{j}-\lfloor\tilde{j}\rfloor\right)\Vec{e}_{y}\times m_{ij}^{*}\Vec{v}_{ij}^{*}}\nonumber\\
\lefteqn{\left.+w_{\lfloor \tilde{i}\rfloor +1\lfloor \tilde{j}\rfloor +1 ij}^{*}
+w_{\lfloor \tilde{i}\rfloor\lfloor \tilde{j}\rfloor +1 ij}^{*}\right)m_{ij}^{*}\Vec{v}_{ij}^{*}}
\hphantom{+\left(\tilde{i}-\lfloor\tilde{i}\rfloor\right)\Vec{e}_{x}\times m_{ij}^{*}\Vec{v}_{ij}^{*}
+\left(\tilde{j}-\lfloor\tilde{j}\rfloor\right)\Vec{e}_{y}\times m_{ij}^{*}\Vec{v}_{ij}^{*}}\nonumber\\
\lefteqn{+\Vec{e}_{x}\times\left(
w_{\lfloor \tilde{i}\rfloor +1\lfloor \tilde{j}\rfloor ij}^{*}+w_{\lfloor \tilde{i}\rfloor +1\lfloor \tilde{j}\rfloor +1 ij}^{*}
\right)m_{ij}^{*}\Vec{v}_{ij}^{*}}
\hphantom{+\left(\tilde{i}-\lfloor\tilde{i}\rfloor\right)\Vec{e}_{x}\times m_{ij}^{*}\Vec{v}_{ij}^{*}
+\left(\tilde{j}-\lfloor\tilde{j}\rfloor\right)\Vec{e}_{y}\times m_{ij}^{*}\Vec{v}_{ij}^{*}}\nonumber\\
\lefteqn{+\Vec{e}_{y}\times\left(
w_{\lfloor \tilde{i}\rfloor +1\lfloor \tilde{j}\rfloor +1 ij}^{*}+w_{\lfloor \tilde{i}\rfloor\lfloor \tilde{j}\rfloor +1 ij}^{*}
\right)m_{ij}^{*}\Vec{v}_{ij}^{*}}
\hphantom{+\left(\tilde{i}-\lfloor\tilde{i}\rfloor\right)\Vec{e}_{x}\times m_{ij}^{*}\Vec{v}_{ij}^{*}
+\left(\tilde{j}-\lfloor\tilde{j}\rfloor\right)\Vec{e}_{y}\times m_{ij}^{*}\Vec{v}_{ij}^{*}}\nonumber\\
\lefteqn{-\sum_{k=\lfloor\tilde{i}\rfloor}^{\lfloor\tilde{i}\rfloor+1}\sum_{l=\lfloor\tilde{j}\rfloor}^{\lfloor\tilde{j}\rfloor+1}
\Vec{r}_{G}^{*}\times w_{klij}^{*}m_{ij}^{*}\Vec{v}_{ij}^{*}}
\hphantom{+\left(\tilde{i}-\lfloor\tilde{i}\rfloor\right)\Vec{e}_{x}\times m_{ij}^{*}\Vec{v}_{ij}^{*}
+\left(\tilde{j}-\lfloor\tilde{j}\rfloor\right)\Vec{e}_{y}\times m_{ij}^{*}\Vec{v}_{ij}^{*}}\nonumber\\
\lefteqn{=\left(\lfloor\tilde{i}\rfloor\Vec{e}_{x}+\lfloor\tilde{j}\rfloor\Vec{e}_{y}\right)\times m_{ij}^{*}\Vec{v}_{ij}^{*}}
\hphantom{+\left(\tilde{i}-\lfloor\tilde{i}\rfloor\right)\Vec{e}_{x}\times m_{ij}^{*}\Vec{v}_{ij}^{*}
+\left(\tilde{j}-\lfloor\tilde{j}\rfloor\right)\Vec{e}_{y}\times m_{ij}^{*}\Vec{v}_{ij}^{*}}\nonumber\\
+\left(\tilde{i}-\lfloor\tilde{i}\rfloor\right)\Vec{e}_{x}\times m_{ij}^{*}\Vec{v}_{ij}^{*}
+\left(\tilde{j}-\lfloor\tilde{j}\rfloor\right)\Vec{e}_{y}\times m_{ij}^{*}\Vec{v}_{ij}^{*}\nonumber\\
\lefteqn{-\Vec{r}_{G}^{*}\times m_{ij}^{*}\Vec{v}_{ij}^{*}}
\hphantom{+\left(\tilde{i}-\lfloor\tilde{i}\rfloor\right)\Vec{e}_{x}\times m_{ij}^{*}\Vec{v}_{ij}^{*}
+\left(\tilde{j}-\lfloor\tilde{j}\rfloor\right)\Vec{e}_{y}\times m_{ij}^{*}\Vec{v}_{ij}^{*}}\nonumber\\
\lefteqn{=\left(\tilde{i}\Vec{e}_{x}+\tilde{j}\Vec{e}_{y}-\Vec{r}_{G}^{*}\right)\times m_{ij}^{*}\Vec{v}_{ij}^{*}}
\hphantom{+\left(\tilde{i}-\lfloor\tilde{i}\rfloor\right)\Vec{e}_{x}\times m_{ij}^{*}\Vec{v}_{ij}^{*}
+\left(\tilde{j}-\lfloor\tilde{j}\rfloor\right)\Vec{e}_{y}\times m_{ij}^{*}\Vec{v}_{ij}^{*}}\nonumber\\
\lefteqn{=\left(\Vec{r}_{ij}+\Vec{v}_{ij}^{*}\tau_{a}-\Vec{r}_{G}^{*}\right)\times m_{ij}^{*}\Vec{v}_{ij}^{*},}
\hphantom{+\left(\tilde{i}-\lfloor\tilde{i}\rfloor\right)\Vec{e}_{x}\times m_{ij}^{*}\Vec{v}_{ij}^{*}
+\left(\tilde{j}-\lfloor\tilde{j}\rfloor\right)\Vec{e}_{y}\times m_{ij}^{*}\Vec{v}_{ij}^{*}}
\end{eqnarray}
and equals the total angular momentum before allocation.
Thus total angular momentum $\Vec{l}_{ij}^{*}$ of gas particles in the cell at lattice point $ij$
does not change in the procedure $T_{a}$.
These results hold in any cell
and thus total mass $m_{G}^{*}$, momentum $\Vec{p}_{G}^{*}$ and angular momentum $\Vec{l}_{G}^{*}$ of the system
are also conserved in the advection procedure $T_{g}$.


\begin{thebibliography}{10}
\bibitem{Kaneko}K. Kaneko and I. Tsuda, "Complex Systems: Chaos and Beyond", Springer-Verlag, Berlin, Heidelberg, New York (2001).
\bibitem{Kanekos}K. Kaneko ed., "Theory and Applications of Coupled Map Lattices", John Wiley \& Sons, Chichester (1993).
\bibitem{Yanagitab}T. Yanagita, "Coupled map lattice model for boiling", \textit{Phys. Lett. A} \textbf{165} (1992) 405.
\bibitem{Yanagitac}T. Yanagita and K. Kaneko, "Coupled map lattice model for convection", \textit{Phys. Lett. A} \textbf{l75} (1993) 415.
\bibitem{Yanagitad}T. Yanagita and K. Kaneko, "Modeling and characterization of cloud dynamics",
\textit{Phys. Rev. Lett.} \textbf{78} (1997) 4297.
\bibitem{Nishimori}H. Nishimori and N. Ouchi, "Formation of ripple patterns and dunes by wind-blown sand",
\textit{Phys. Rev. Lett.} \textbf{71} (1993) 197.
\bibitem{GalaDy}J. Binney and S. Tremaine, "Galactic Dynamics", second edition, Princeton University Press, New Jersey (2008).
\bibitem{SEEDS}M. Tamura, "SEEDS---Strategic explorations of exoplanets and disks
with the Subaru Telescope---", \textit{Proc. Jpn. Acad., Ser. B} \textbf{92} (2016) 45.
\bibitem{Lin-Shu}C.C. Lin and F.H. Shu, "On the spiral structure of disk galaxies", \textit{ApJ} \textbf{140} (1964) 646.
\bibitem{Kuno}N. Kuno, \textit{et al.}, "Two-arm spiral structure of molecular gas in the flocculent galaxy NGC 5055",
\textit{Publ. Astron. Soc. Japan} \textbf{49} (1997) 275.
\bibitem{Miyoshi}M. Miyoshi, \textit{et al.}, "Evidence for a black hole from high rotation velocities in a sub-parsec region of NGC4258", \textit{Nature} \textbf{373} (1995) 127.
\bibitem{Nadia}N.M. Murillo, \textit{et al.}, "A Keplerian disk around a Class 0 source: ALMA observations of VLA1623A", \textit{A\&A} \textbf{560}, A103 (2013).
\bibitem{Kuno2}N. Kuno and N. Nakai, "Distribution and dynamics of molecular gas in the galaxy M51. III. Kinematics of molecular gas",
\textit{Publ. Astron. Soc. Japan} \textbf{49} (1997) 279.
\bibitem{Tombesi}F. Tombesi, \textit{et al.}, "Evidence for ultrafast outflows in radio-quiet AGNs --- III. Location and energetics", \textit{Mon. Not. R. Astron. Soc.} \textbf{422} (2012) L1.
\bibitem{Matsushita}Y. Matsushita, \textit{et al.}, "A very compact extremely high velocity flow toward MMS 5/OMC-3 revealed with ALMA", \textit{ApJ} \textbf{871} (2019) 221.
\bibitem{Matsumoto}T. Matsumoto and T. Hanawa,
"Fragmentation of a molecular cloud core versus fragmentation of the massive protoplanetary disk in the main accretion phase",
\textit{ApJ} \textbf{595} (2003) 913.
\bibitem{AMRo}M.J. Berger and P. Colella, "Local adaptive mesh refinement for shock hydrodynamics",
\textit{J. Comput. Phys.} \textbf{82} (1989) 64.
\bibitem{SPHo}R.A. Gingold and J.J. Monaghan, "Smoothed particle hydrodynamics: theory and application to non-spherical stars",
\textit{Mon. Not. R. Astron. Soc.} \textbf{181} (1977) 375.
\bibitem{Springel}V. Springel and L. Hernquist, "Cosmological SPH simulations: a hybrid multi-phase model for star formation",
\textit{Mon. Not. R. Astron. Soc.} \textbf{339} (2003) 289.
\bibitem{Tsuda}K. Kaneko and I. Tsuda, "Chaotic itinerancy", \textit{Chaos} \textbf{13} (2003) 926.
\bibitem{Nozawa}H. Nozawa, "A neural network model as a globally coupled map and applications based on chaos",
\textit{Chaos} \textbf{2} (1992) 377.
\bibitem{Konishi}T. Tsuchiya, N. Gouda and T. Konishi,
"Chaotic Itinerancy and Thermalization in a One-Dimensional Self-Gravitating System",
\textit{Astrophys. Space Sci.} \textbf{257} (1997) 319.
\bibitem{Faber}S.M. Faber, \textit{et al.},
"Galaxy luminosity functions to z$\sim$1 from DEEP2 and COMBO-17: implications for red galaxy formation",
\textit{ApJ} \textbf{665} (2007) 265.
\bibitem{Koyama}S. Koyama, \textit{et al.},
"Do galaxy morphologies really affect the efficiency of star formation during the phase of galaxy transition?",
\textit{ApJ} \textbf{874} (2019) 142.
\end{thebibliography}
\end{document}